\documentclass[apj]{emulateapj} 

\usepackage{epsfig}
\usepackage{amsmath}
\usepackage{amssymb}
\usepackage{natbib}
\usepackage{threeparttable} 
\usepackage{booktabs}

\usepackage{epstopdf}

\newcommand{\chan}{\textit{Chandra}}
\newcommand{\swift}{\textit{Swift}}
\newcommand{\rxte}{\textit{RXTE}}
\newcommand{\xmm}{\textit{XMM-Newton}}
\newcommand{\inte}{\textit{INTEGRAL}}
\newcommand{\maxi}{\textit{MAXI}}

\newcommand{\suzaku}{\textit{Suzaku}}
\newcommand{\beppo}{\textit{BeppoSAX}}
\newcommand{\fermi}{\textit{Fermi}}

\newcommand{\Msun}{\mathrm{M}_{\odot}}
\newcommand{\lum}{\mathrm{erg~s}^{-1}}
\newcommand{\flux}{\mathrm{erg~cm}^{-2}~\mathrm{s}^{-1}}
\newcommand{\fluence}{\mathrm{erg~cm}^{-2}}
\newcommand{\cnts}{\mathrm{counts~s}^{-1}}
\newcommand{\mdot}{\mathrm{M_{\odot}~yr}^{-1}}
\newcommand{\nh}{\mathrm{cm}^{-2}}

\newcommand{\new}{Swift J185003.2--005627}
\newcommand{\source}{Swift J1922.7--1716}

\def \aj {AJ}
\def \mnras {MNRAS}
\def \apj {ApJ}
\def \apjs {ApJS}
\def \apjl {ApJL}
\def \aap {A\&A}

\def \araa {ARA\&A}

\def \pasj {PASJ}

\def \gca {Geochim. Cosmochim. Acta}

\slugcomment{Published in The Astrophysical Journal, 759, 8 (2012)}
\shorttitle{Two new bursting neutron stars}
\shortauthors{Degenaar et al.}

\begin{document}

\title{Two new bursting neutron star low-mass X-ray binaries:\\ Swift J185003.2--005627 and Swift J1922.7--1716}

\author{N. Degenaar\altaffilmark{1}}
\affil{Department of Astronomy, University of Michigan, 500 Church St, Ann Arbor, MI 48109, USA}
\email{degenaar@umich.edu}

\author{M. Linares}
\affil{Massachusetts Institute of Technology (MIT), Kavli Institute for Astrophysics and Space Research, Cambridge, MA 02139, USA}

\author{D. Altamirano, and R. Wijnands}
\affil{Astronomical Institute "Anton Pannekoek", University of Amsterdam, Postbus 94249, 1090 GE Amsterdam, The Netherlands}

\altaffiltext{1}{Hubble fellow}


\begin{abstract} 
We discuss the origin of two triggers of \swift's Burst Alert Telescope (BAT) that occurred in 2011. The triggers were identified with \new, a previously unknown X-ray source, and the known but unclassified X-ray transient \source. We investigate the BAT data and follow-up observations obtained with \swift's X-ray and ultraviolet/optical telescopes to demonstrate that both triggers are consistent with thermonuclear X-ray bursts. This implies that both sources are neutron star low-mass X-ray binaries. The total duration of $\simeq7$~minutes and estimated energy output of $\simeq(3-7)\times10^{39}$~erg, fall in between that of normal and intermediately long X-ray bursts. From the observed peaks of the X-ray bursts, we estimate a distance of $\lesssim 3.7$~kpc for \new\ and $\lesssim 4.8$~kpc for \source. 
We characterize the outburst and quiescent X-ray properties of the two sources. They have comparable average outburst luminosities of $\simeq 10^{35-36}~\lum$, and a quiescent luminosity equal to or lower than $\simeq2\times 10^{32}~\lum$ (0.5--10 keV). \new\ returned to quiescence $\simeq20$~d after its BAT trigger, while \source\ appears to exhibit long accretion outbursts that last several months to years. We identify a unique counterpart for \source\ in the ultraviolet/optical data. Finally, we serendipitously detect a flare lasting $\simeq500$~s from an uncataloged X-ray/optical object that we tentatively classify as a flaring M-dwarf.\\
\end{abstract}



\section{Introduction}\label{sec:intro}
Low-mass X-ray binaries (LMXBs) consist of a neutron star or a black hole that accretes matter from a (sub-) solar mass companion star that overflows its Roche lobe. The accretion process involves the formation of an accretion disk and typically generates a 0.5--10 keV X-ray luminosity of $L_{\mathrm{X}}\simeq10^{36-39}~\lum$ \citep[e.g.,][]{chen97}, which places LMXBs amongst the brightest X-ray point sources in our Galaxy. 
Many LMXBs have an unstable accretion disk \citep[see][for a review]{lasota01}. This causes the system to become \textit{transient}, i.e., to alternate accretion outbursts with long episodes of quiescence during which the X-ray emission is much dimmer \citep[$ \lesssim10^{33}~\lum$; e.g.,][]{menou99,garcia01}. Typically, the active phases of transient LMXBs last for a few weeks or months, while they reside in quiescence for years before a new outburst commences.

Whereas it is not straightforward to distinguish the nature of the compact object from the X-ray spectral and timing properties of the outburst \citep[e.g.,][]{vanderklis1994,linares2007}, there are two phenomena that require a solid surface and are therefore considered a distinctive property of accreting neutron stars. These are coherent X-ray pulsations and thermonuclear X-ray bursts (i.e., type-I X-ray bursts, X-ray bursts hereafter). X-ray bursts are bright flashes of X-ray emission that may reach up to the Eddington limit and temporarily outshine the accretion luminosity. They are caused by unstable burning of accreted helium (He) and/or hydrogen (H) on the surface of the neutron star.

X-ray bursts are characterized by blackbody emission that peaks at a temperature up to $kT_{\mathrm{bb}}\simeq2-3$~keV and decays to $\simeq 1$~keV as the X-ray burst fades. 
The observable properties such as the duration ($t_b$), radiated energy ($E_b$), and recurrence time ($t_{\mathrm{rec}}$) depend on the amount and composition of the fuel that is consumed during the X-ray burst. 
The most commonly observed X-ray bursts last for $t_b \simeq10-100$~s, have a radiated energy output of $E_b \simeq 10^{39}$~erg, and recur on a timescale of minutes to days. To date, thousands of such events have been observed from about $\simeq100$ different LMXBs \citep[see, e.g.,][]{cornelisse2003,galloway06}. 

Apart from these {\it normal} X-ray bursts, there are two classes of more energetic events that are also more rare: intermediately long X-ray bursts \citep[$t_b\simeq$ tens of minutes to hours, $E_b\simeq10^{40-41}$~erg, and $t_{\mathrm{rec}}\simeq$ weeks to years; see][]{chenevez2008,falanga08} and superbursts \citep[$t_b\simeq$ hours to days, $E_b\simeq10^{42}$~erg, and $t_{\mathrm{rec}}\simeq$ a year; see][]{kuulkers2004,keek2008}. To date, only a few tens of such energetic X-ray bursts have been observed from about two dozens of LMXBs.

In this work, we discuss the properties of \new\ and \source. We demonstrate that both sources displayed an X-ray burst in 2011 that triggered \swift's Burst Alert Telescope \citep[BAT;][]{barthelmy05}. We use rapid follow-up observations obtained with the narrow-field X-Ray Telescope \citep[XRT;][]{burrows05} and Ultra-Violet/Optical Telescope \citep[UVOT;][]{roming05} to fully characterize the trigger events. In addition, we investigate the outburst and the quiescent properties of both sources.


\subsection{The New X-Ray Source \new}\label{subsec:new_intro}
On 2011 June 24, \swift/BAT triggered and located an unknown source \citep[trigger 456014;][]{beardmore2011_gcn}. Based on the soft nature of the BAT detection and the proximity to the Galactic plane, the trigger was tentatively identified as a previously unknown Galactic transient and dubbed \new\ (J1850 hereafter). The detection of a fading X-ray counterpart in follow-up XRT observations provided an arcsecond localization of the source \citep[][]{beardmore2011_gcn}. Based on the properties of the XRT data and the refined analysis of the BAT trigger data\footnote{See also http://gcn.gsfc.nasa.gov/notices$\_$s/456014/BA/}, it was suggested that this event was a thermonuclear X-ray burst \citep[][]{markwardt2011,beardmore2011_atel}.

J1850 was not detected in the \swift/BAT hard X-ray transient monitor on or after the time of the BAT trigger, with a 1-$\sigma$ upper limit on the daily averaged 15--50 keV count rate of $7\times10^{-4}~\cnts~\nh$ \citep[][]{krimm2011}. However, the source was detected at an average 15--50 keV intensity of $(3.5\pm1.0)\times10^{-3}~\cnts~\nh$ (corresponding to a flux of $\simeq2\times10^{-10}~\flux$ for a Crab-like spectrum) between 2011 May 18 and May 26. Archival searches back to 2005 February did not reveal any similar detections of J1850 \citep[][]{krimm2011}.

\begin{table}
\caption{Log of (Soft) X-ray Observations.}
\begin{center}
\begin{threeparttable}
\begin{tabular}{c c c c c}
\toprule
Satellite/Instrument & Obs ID & Date & $t_{\mathrm{exp}}$ (ks) \\
\midrule
\multicolumn{4}{c}{{\bf \new}}  \\
\xmm/EPIC* & 73740101 & 2003 Mar 21 & 26.8  \\
\xmm/EPIC* & 73740201 & 2003 Mar 23 & 20.5   \\
\xmm/EPIC* & 73740301 & 2003 Apr 18 & 24.2  \\   
\swift/XRT (PC+WT) & 456014000 & 2011 Jun 24 & 1.8	 \\
\swift/XRT (PC+WT) & 456014001 & 2011 Jun 25 & 1.0	 \\	
\swift/XRT (PC+WT) & 456014002 & 2011 Jun 25 & 1.5	 \\	
\swift/XRT (PC+WT) & 456014003 & 2011 Jun 25 & 1.5	 \\	
\swift/XRT (WT) & 456014004 & 2011 Jun 26 & 1.0	 \\	
\swift/XRT (WT) & 456014005 & 2011 Jun 27 & 1.5	 \\	
\swift/XRT (WT) & 456014007 & 2011 Jun 28 & 4.9	 \\	
\swift/XRT (WT) & 456014008 & 2011 Jun 29 & 1.0	 \\	
\swift/XRT (WT) & 456014010 & 2011 Jul 1 & 1.0	 \\
\swift/XRT (WT) & 456014011 & 2011 Jul 2 & 0.4	 \\
\swift/XRT (WT) & 456014012 & 2011 Jul 3 & 1.2	 \\
\swift/XRT (WT) & 456014013 & 2011 Jul 4 & 1.0 	 \\	
\swift/XRT (WT) & 456014014 & 2011 Jul 5 & 0.5	 \\
\swift/XRT (WT) & 456014015 & 2011 Jul 6 & 1.0  \\	
\swift/XRT (WT) & 456014016 & 2011 Jul 7 & 1.0 	 \\	
\swift/XRT (WT) & 456014017 & 2011 Jul 8 & 0.9 \\	
\swift/XRT (WT) & 456014019 & 2011 Jul 10 & 0.5  \\	
\swift/XRT (PC) & 456014021 & 2011 Jul 12 & 1.1  \\	
\swift/XRT (PC) & 456014022 & 2011 Jul 13 & 1.0	 \\	
\swift/XRT (PC)* & 456014023 & 2011 Jul 16 & 0.2	 \\	
\swift/XRT (PC)* & 456014024 & 2011 Jul 17 & 0.8	 \\	
\swift/XRT (PC)* & 456014025 & 2011 Jul 18 & 0.8  \\	
\swift/XRT (PC)* & 456014026 & 2011 Jul 19 & 0.4  \\	
\swift/XRT (PC)* & 456014027 & 2011 Jul 20 & 0.1  \\	
\swift/XRT (PC)* & 456014028 & 2011 Jul 21 & 1.0  \\       
\midrule
\multicolumn{4}{c}{{\bf \source}}  \\
\swift/XRT (PC+WT)$\dagger$ & 35174001 & 2005 Jul 8 & 6.4 \\ 
\swift/XRT (PC+WT)$\dagger$ & 35174003 & 2005 Oct 1 & 10.8  \\ 
\swift/XRT (PC+WT) & 35471001 & 2006 Mar 14 & 16.2  \\ 
\swift/XRT (PC+WT) & 35471002 & 2006 Jun 4 & 0.3  \\ 
\swift/XRT (PC+WT) & 35471003 & 2006 Jun 18 & 10.8  \\ 
\swift/XRT (PC)* & 35471004 & 2006 Oct 30 & 5.2 \\ 
\swift/XRT (PC)* & 35471005 & 2006 Nov 3 & 1.1 \\ 
\swift/XRT (PC)* & 35471006 & 2006 Nov 10 & 4.4 \\ 
\swift/XRT (PC)* & 35471007 & 2006 Nov 14 & 3.7 \\ 
\suzaku/XIS* & 702028010 & 2007 Apr 10 & 78.6 \\ 
\swift/XRT (PC+WT) & 35471008 & 2011 Aug 13 & 1.0 \\ 
\swift/XRT (PC+WT) & 35471009 & 2011 Aug 30 & 1.2 \\ 
\swift/XRT (PC+WT) & 35471010 & 2011 Aug 31 & 4.2 \\ 
\swift/XRT (WT) & 506913000 & 2011 Nov 3 & 1.9 \\ 
\swift/XRT (WT) & 35471011 & 2011 Nov 4 & 0.9 \\ 
\swift/XRT (WT) & 35471012 & 2011 Nov 5 & 1.0 \\ 
\swift/XRT (PC+WT) & 35471013 & 2012 Mar 9 & 0.7 \\ 
\swift/XRT (PC+WT) & 35471014 & 2012 Mar 14 & 1.0 \\ 
\swift/XRT (PC)* & 35471015 & 2012 May 25 & 1.2 \\ 
\swift/XRT (PC)* & 35471016 & 2012 Jun 9 & 1.9 \\ 
\swift/XRT (PC)* & 35471017 & 2012 Jun 11 & 2.0 \\ 
\swift/XRT (PC)* & 35471018 & 2012 Jun 13 & 0.9 \\ 
\swift/XRT (PC)* & 35471019 & 2012 Jun 15 & 0.7 \\ 
\swift/XRT (PC)* & 35471020 & 2012 Jun 17 & 0.3 \\ 
\swift/XRT (PC)* & 35471021 & 2012 Jun 21 & 2.3 \\ 
\bottomrule
\end{tabular}
\label{tab:obs}
\begin{tablenotes}
\item[]Note. -- Observations marked by an asterisk were obtained during quiescent episodes. The two observations marked by a dagger were also discussed by \citet{falanga2006}.
\end{tablenotes}
\end{threeparttable}
\end{center}
\end{table}


\subsection{The Unclassified Source \source}\label{subsec:source_intro}
\source\ (hereafter J1922) is a transient X-ray source that was discovered during the \swift/BAT hard X-ray survey of 2004 December -- 2005 March \citep[][]{tueller2005a,tueller2005b}. \swift/XRT follow-up observations allowed for the identification of a soft X-ray counterpart \citep[][]{tueller2005b}. Non-detections with \inte\ \citep[2003--2004;][]{falanga2006} and \swift\ \citep[2006 October--November;][]{kennea2011_bursters} testified to its transient nature. 

\citet{falanga2006} discussed \swift, \rxte, and \inte\ data taken between 2005 July and November. The broadband spectrum and timing properties were found to be typical of an LMXB in a faint X-ray state, but the nature of the compact accretor (i.e., a neutron star or a black hole) could not be established. A combined blackbody and power-law model provided an adequate description of the broadband spectrum, with a hydrogen column density $N_{\mathrm{H}} \simeq (1.5-2.0)\times10^{21}~\nh$, spectral index $\Gamma \simeq 1.4-1.8$, and temperature $kT_{\mathrm{bb}} \simeq 0.4-0.6$~keV. The resulting 0.1--100 keV flux was $\simeq (3.3-3.9)\times10^{-10}~\flux$ \citep[][]{falanga2006}.

Renewed activity of J1922 was seen with \maxi\ and \swift/XRT in 2011 August \citep[][]{nakahira2011,kennea2011_bursters}. Inspection of the \swift/BAT transient monitor data revealed that the source was detected in hard X-rays starting in 2011 July. It was seen at a mean 15--50 keV count rate of $(1.5 \pm 0.3)\times10^{-3}~\cnts~\mathrm{cm}^{-2}$ ($\simeq1\times10^{-10}~\flux$ for a Crab-like spectrum) till early-August, after which the hard X-ray flux decreased \citep[][]{kennea2011_bursters}. The 2011 intensity was similar to the average BAT count rate of the source in 2005.

\swift/BAT triggered on J1922 on 2011 November 3 \citep[][]{barthelmy2011}.\footnote{See also http://gcn.gsfc.nasa.gov/notices$\_$s/506913/BA/} 
Preliminary analysis of the BAT and XRT data revealed that the trigger was very likely caused by an X-ray burst \citep[][]{degenaar2011_j1922}. Optical spectroscopy revealed clear He and H emission lines, supporting an LMXB nature \citep[][]{wiersema2011,halpern2011}. The presence of H-lines implies that the donor star is H-rich and therefore the binary orbital period must be $P_{\mathrm{orb}}\gtrsim 1.5$ hr \citep[][]{nelson86}. Upper limits on the quiescent optical counterpart ($>23.4$~mag in the $r$ and $g$ bands) are consistent with an M-dwarf or evolved star and suggest $P_{\mathrm{orb}}\lesssim5$~hr \citep[][]{halpern2011}.


\section{Observations and data analysis}\label{sec:obs}
We use publicly available \swift\ data to investigate the BAT triggers and outburst properties of J1850 and J1922. In addition, we analyze archival \xmm\ and \suzaku\ data in an attempt to constrain their quiescent emission level. Table~\ref{tab:obs} gives an overview of all observations discussed in this work. The observations, data reduction and analysis procedures are detailed in the following sections. All errors quoted in this work refer to 90\% confidence levels unless stated otherwise.


\subsection{\swift/BAT}\label{subsec:bat}
We generated standard BAT data products (15--150 keV) for the trigger observations using the \textsc{batgrbproduct} tool. In all cases, the spacecraft started slewing towards the trigger location when the source had already faded into the background. We extracted single BAT spectra of the pre-slew data employing the tool \textsc{batbinevt}. The standard geometrical corrections and the BAT-recommended systematical error were applied with \textsc{batupdatephakw} and \textsc{batphasyserr}, respectively. Since we use only pre-slew data, we generated a single response matrix for each trigger by running the task \textsc{batdrmgen}. The spectra were fitted between 15 and 35 keV with \textsc{XSpec} \citep[ver. 12.7;][]{xspec}. 


\subsection{\swift/XRT}\label{subsec:xrt}
The XRT data cover an energy range of 0.5--10 keV and consist of a combination of photon counting (PC) and windowed timing (WT) mode. In the PC mode, a two-dimensional image is acquired, whereas in the WT mode the CCD columns are collapsed into a one-dimensional image to reduce the frame time. The WT is typically used when the count rate exceeds $\simeq1-2~\cnts$, because higher rates cause considerable pile-up in the PC mode. 

All {\it Swift} data were reduced using the \swift\ tools (ver. 3.8) within the \textsc{heasoft} package (v. 6.12), and employing the latest calibration data (ver. 3.8). We made use of the online tools to obtain XRT data products and obtain a global characterization of the persistent emission \citep[][]{evans09}.\footnote{http://www.swift.ac.uk/user$\_$objects/} For the detailed analysis of the X-ray bursts we manually extracted XRT light curves and spectra using \textsc{XSelect} (ver. 3.8). Exposure maps were generated using the tool \textsc{xrtexpomap}, and subsequently used to create ancillary response files (arfs) with \textsc{xrtmakearf}. The latest redistribution matrix files (rmfs) were taken from the calibration database (\textsc{caldb}).

The XRT spectra were grouped to contain a minimum of 20 photons per bin and fitted between 0.5 and 10 keV in \textsc{XSpec}.


\subsection{\swift/UVOT}\label{subsec:uvot}
The UVOT data were obtained using a variety of optical and ultraviolet (UV) filters in a wavelength range of $\simeq 1500-8500$~\AA{} \citep[see][]{poole2008}. We used a standard aperture of $5''$ to extract source photons and a source-free region with a radius of $10''$ as a background reference. Magnitudes and light curves were extracted using the tools \textsc{uvotsource} and \textsc{uvotmaghist}.


\subsection{Spectral Analysis and Eddington Limit}\label{subsec:ana}
For our spectral analysis with \textsc{XSpec}, we use a blackbody model (BBODYRAD), a simple power law (POWERLAW), or a combination of both. In all fits, we included the PHABS model to account for neutral hydrogen absorption along the line of sight, for which we employed the default \textsc{XSpec} abundances \citep[][]{anders1989_phabs_abun} and cross-sections \citep[][]{balucinska1992_phabs_cross}. 

The bolometric accretion luminosity is typically a factor $\simeq2-3$ higher than observed in the 0.5--10 keV energy band \citep[][]{zand07}. We therefore apply a correction factor of $2.5$ to the 0.5--10 keV luminosity ($L_{\mathrm{X}}$, determined from the \swift/XRT data) to estimate the bolometric accretion luminosity ($L_{\mathrm{bol}}$). We subsequently use this to estimate the mass accretion rate onto the neutron star (during outburst) via the relation $\dot{M}_{\mathrm{ob}}=RL_{\mathrm{bol}}/GM$, where $G=6.67\times10^{-8}~\mathrm{cm}^{3}~\mathrm{g}^{-1}~\mathrm{s}^{-2}$ is the gravitational constant, and $M$ and $R$ are the mass and radius of the neutron star, respectively. We adopt canonical values of $M=1.4~\Msun$ and $R=10$~km. 

It is often useful to express the accretion luminosity and mass-accretion rate in terms of the Eddington limit. We will adopt $L_{\mathrm{EDD}}=3.8\times10^{38}~\lum$, which is the empirical limit determined from X-ray bursts that display photospheric radius expansion \citep[PRE;][]{kuulkers2003}.

\begin{table*}
\caption{Results from Spectral Fitting of the Averaged Outburst Data.}
\begin{center}
\begin{threeparttable}
\begin{tabular}{c c c c c c c c c}
\toprule
Year & $N_{\mathrm{H}}$ & $\Gamma$ & $kT_{\mathrm{bb}}$ & $R_{\mathrm{bb}}$ & $\chi_{\nu}^2$ (dof) & $F_{\mathrm{X}}$ & $L_{\mathrm{X}}$ & $L_{\mathrm{bol}}$ \\
Data Mode & ($10^{22}~\nh$) &  & (keV) & (km) & & ($10^{-10}~\flux$) & \multicolumn{2}{c}{($10^{35}~\lum$)} \\
\midrule
\multicolumn{9}{c}{{\bf \new}} \\
2011 & $1.1\pm 0.1$ & \nodata  & \nodata & \nodata & 1.1 (669) & \nodata & \nodata & $\simeq7$ \\
WT & \nodata & $1.6\pm 0.2$ & $0.7\pm 0.1$ & $1.8\pm 0.2$ & \nodata & $2.3\pm 0.2$ & $3.8\pm0.3$ & \nodata \\
PC & \nodata & $0.9\pm 0.5$ & $0.7\pm 0.1$ & $1.7\pm 0.3$ & \nodata & $1.1\pm 0.1$ & $1.8\pm 0.2$ & \nodata \\
\midrule
\multicolumn{9}{c}{{\bf \source}} \\ 
2005--2006 & $0.17\pm0.08$ & \nodata & \nodata & \nodata & 0.9 (209) & \nodata & \nodata & $\simeq 17$ \\
WT & \nodata & $1.5\pm 0.4$ & $0.4\pm 0.1$ & $6.6\pm 0.7$ & \nodata & $2.8\pm 0.3$ & $7.7\pm 0.8$ & \nodata \\
PC & \nodata & $1.6\pm 0.4$ & $0.4\pm 0.1$ & $5.8\pm 0.7$ & \nodata & $2.2\pm 0.3$ & $6.0\pm 0.9$ & \nodata\\
2011--2012 & $0.15\pm0.03$ & \nodata & \nodata & \nodata & 1.0 (660) & \nodata & \nodata & $\simeq 27$ \\
WT & \nodata & $1.7\pm 0.1$ & $0.7\pm 0.1$ & $3.8\pm 0.4$ & \nodata & $6.1\pm 0.2$ & $16.8\pm0.5$ & \nodata\\
PC & \nodata & $1.6\pm 0.2$ & $0.5\pm 0.1$ & $4.7\pm 0.4$ & \nodata & $1.7\pm 0.1$ & $4.7\pm 0.3$ & \nodata  \\
\bottomrule
\end{tabular}
\label{tab:spec}
\begin{tablenotes}
\item[]Note. All quoted errors refer to $90\%$ confidence levels. The outburst data were fitted to a combined power-law and blackbody model (POWERLAW+BBODYRAD), modified by absorption (PHABS). $F_{\mathrm{X}}$  gives the total unabsorbed model flux in the 0.5--10 keV band and $L_{X}$ represents the corresponding luminosity assuming distances of $3.7$ and $4.8$~kpc for J1850 and J1922, respectively. $L_{\mathrm{bol}}$ represents the estimated bolometric accretion luminosity averaged over the outburst (see the text).
\end{tablenotes}
\end{threeparttable}
\end{center}
\end{table*}

\begin{table*}
\caption{Spectral Parameters of the X-Ray Bursts.}
\begin{center}
\begin{threeparttable}
\begin{tabular}{c c c c c c c}
\toprule
Year & $\Delta t$  & $N_{\mathrm{H}}$ & $kT_{\mathrm{bb}}$ & $R_{\mathrm{bb}}$ & $\chi_{\nu}^2$ (dof) & $F_{\mathrm{bol}}$ \\
Data mode & (s) & ($10^{22}~\nh$) & (keV) & (km) & & ($\flux$)  \\
\midrule
\multicolumn{7}{c}{{\bf \new}} \\
\multicolumn{7}{l}{{2011 June 24}} \\ 
BAT & 0--20 & 1.1 fix & $2.3\pm 0.6$ & $5.6^{+6.5}_{-5.6}$ & 1.0 (7) & $(7.5\pm 4.2)\times10^{-8}$ \\
XRT/WT & 101--201 & 1.1 fix & $0.81\pm 0.03$ & $5.4\pm 0.1$ & 1.1 (66) & $(1.0\pm0.1)\times10^{-9}$ \\
XRT/WT & 202--430 & 1.1 fix & $0.73\pm 0.03$ & $4.4\pm 0.1$ & 1.2 (68) & $(4.5\pm0.1)\times10^{-10}$ \\
\midrule
\multicolumn{7}{c}{{\bf \source}} \\ 
\multicolumn{7}{l}{{2011 November 3}} \\ 
BAT & 0--20 & 0.16 fix & $2.4\pm 0.5$ & $6.7^{+4.0}_{-6.7}$ & 1.7 (8) & $(6.8\pm3.2)\times10^{-8}$ \\
XRT/WT & 136--210 & 0.16 fix & $0.80\pm0.07$ & $5.3\pm0.3$ & 0.9 (84) & $(5.0\pm0.3)\times10^{-10}$ \\
XRT/WT & 211--440 & 0.16 fix & $0.59\pm0.09$ & $5.0\pm0.6$ & 1.2 (151) & $(1.4\pm 0.1)\times10^{-10}$ \\
\multicolumn{7}{l}{{2011 December 2}} \\ 
BAT & 0--20 & 0.16 fix & $2.0\pm 0.5$ & $14.9^{+10.7}_{-14.9}$ & 1.0 (9) & $(10.2\pm6.3)\times10^{-8}$ \\
\bottomrule
\end{tabular}
\label{tab:burst_spec}
\begin{tablenotes}
\item[]Note. All quoted errors refer to $90\%$ confidence levels. The parameter $\Delta t$ indicates the time since the BAT trigger. The burst data were fitted to a blackbody model (BBODYRAD), modified by absorption (PHABS), with the hydrogen column density ($N_{\mathrm{H}}$)  fixed to the value obtained for the outburst fits. When calculating the emitting radii ($R_{\mathrm{bb}}$) we assumed distances of $3.7$ and $4.8$~kpc for J1850 and J1922, respectively. $F_{\mathrm{bol}}$ gives an estimate of the bolometric flux, which was obtained by extrapolating the blackbody fits to the 0.01--100 keV energy range. We note that the emitting radii inferred from simple blackbody fits are expected to be underestimated due to the fact that electron scatterings in the neutron star atmosphere harden the spectrum resulting in a color temperature that is larger than the effective temperature by a factor $\simeq1.5$ \citep[e.g.,][]{suleimanov2011}. Consequently, the inferred radius is underestimated by that same factor.
\end{tablenotes}
\end{threeparttable}
\end{center}
\end{table*}


\section{Results for Swift J185003.2--005627}\label{subsec:new_results}


\begin{figure*}
 \begin{center}
\includegraphics[width=12.0cm]{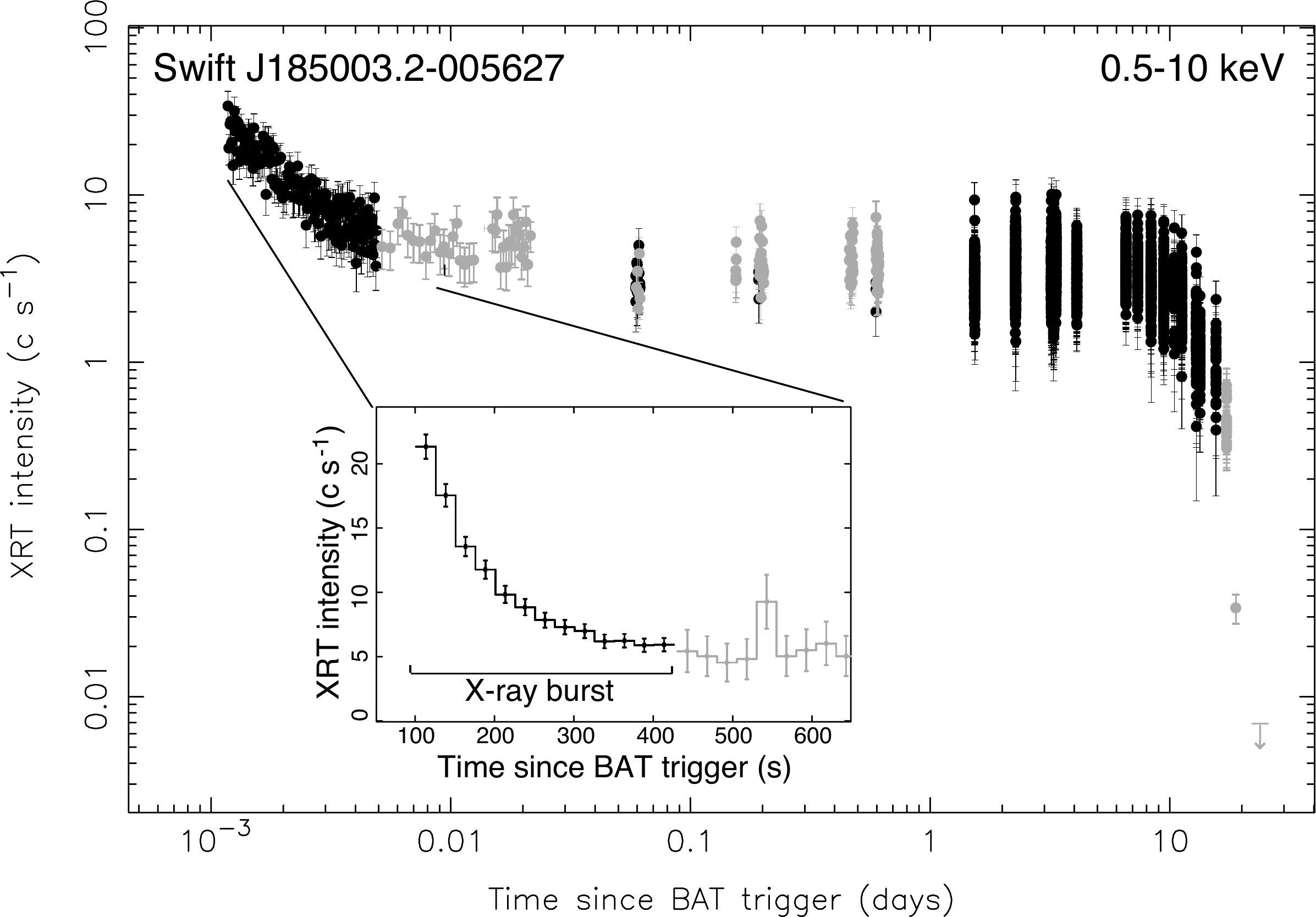}
    \end{center}
\caption[]{{\swift/XRT 0.5--10 keV count rate light curves of J1850 using both WT (black) and PC (gray) data. The BAT trigger (\# 456041) occurred on 2011 June 24. The main image is a log-log plot showing the X-ray burst and subsequent outburst evolution using 10 counts per bin. The inset displays the X-ray burst light curve at 25 s resolution on a linear scale.}}
 \label{fig:lc_1850}
\end{figure*}

\begin{figure}
 \begin{center}
\includegraphics[width=8.0cm]{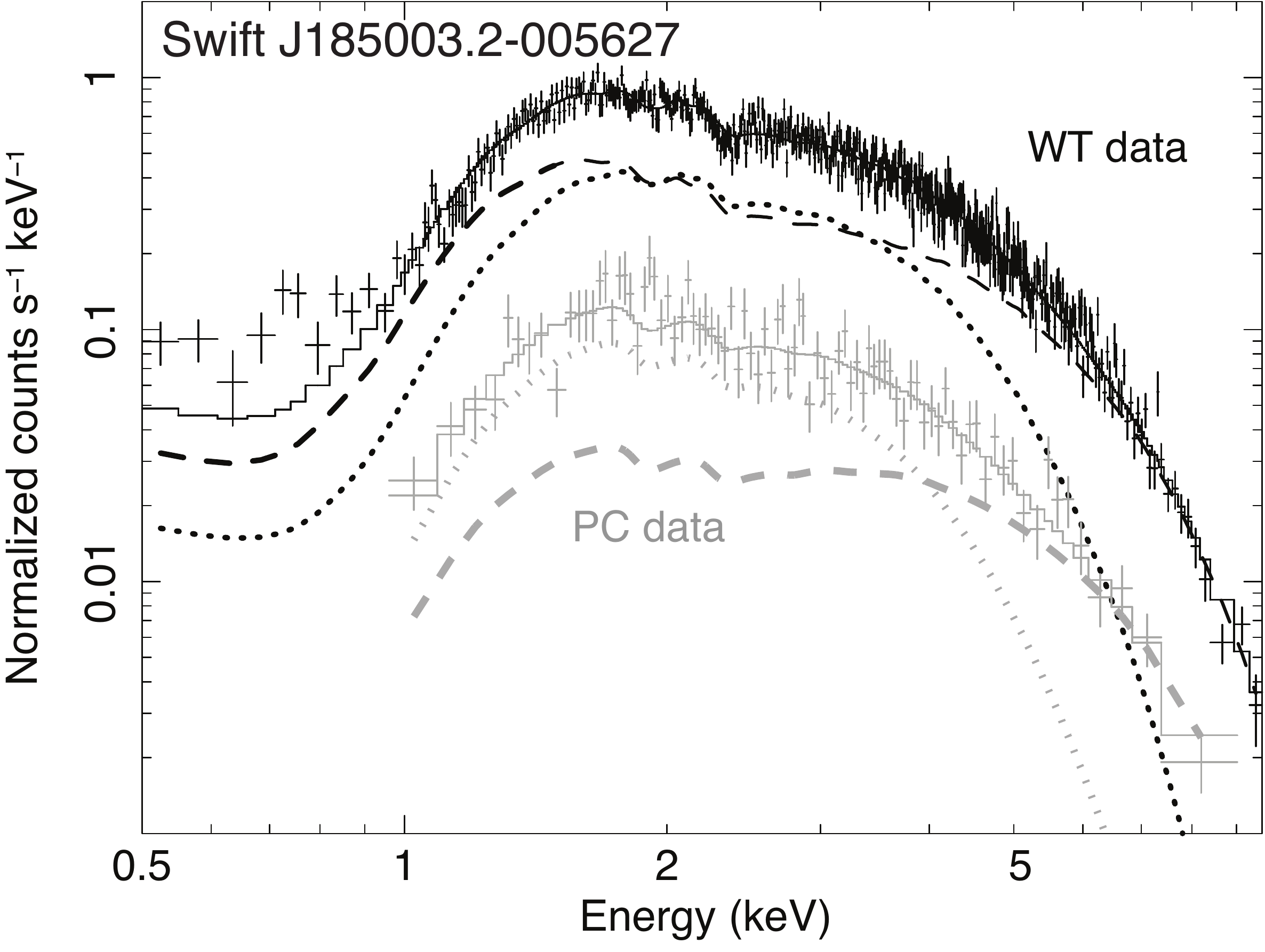}
    \end{center}
\caption[]{{\swift/XRT spectra of the 2011 outburst of J1850, showing both WT (black) and PC (gray) data. The solid lines represent the best fit to a combined power law (dashed line) and blackbody (dotted curve) model. 
}}
 \label{fig:spec}
\end{figure}

\begin{figure}
 \begin{center}
\includegraphics[width=8.0cm]{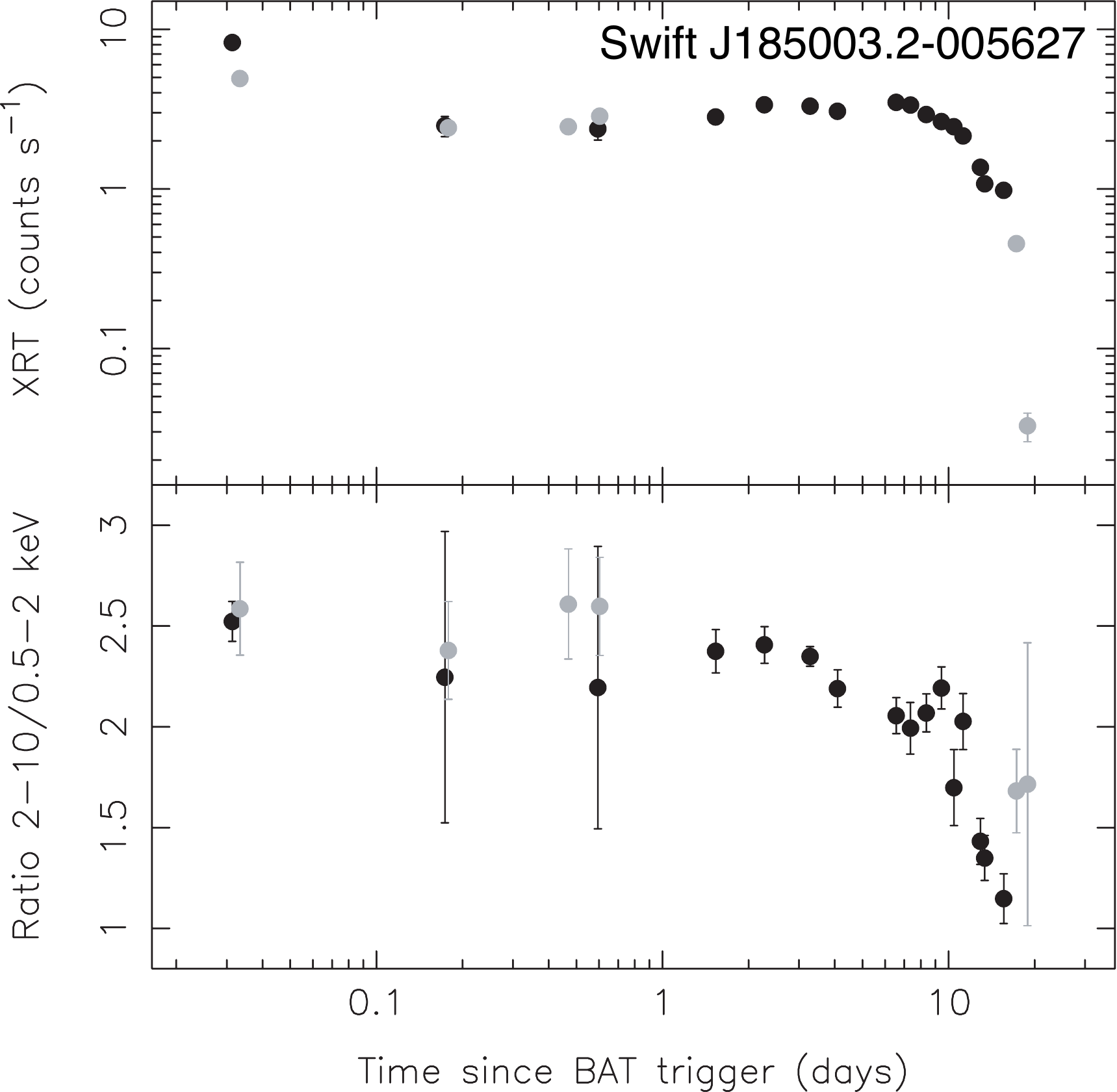}
    \end{center}
\caption[]{{Evolution of the 2011 outburst of J1850 as seen with \swift/XRT. The black and gray points represent WT and PC mode data, respectively. The top plot shows the intensity in the full (0.5--10 keV) energy band, whereas the bottom plot shows the ratio of counts in the hard (2--10 keV) and soft (0.5--2 keV) bands.
}}
 \label{fig:hardness}
\end{figure}

\begin{figure*}
 \begin{center}
\includegraphics[width=5.8cm]{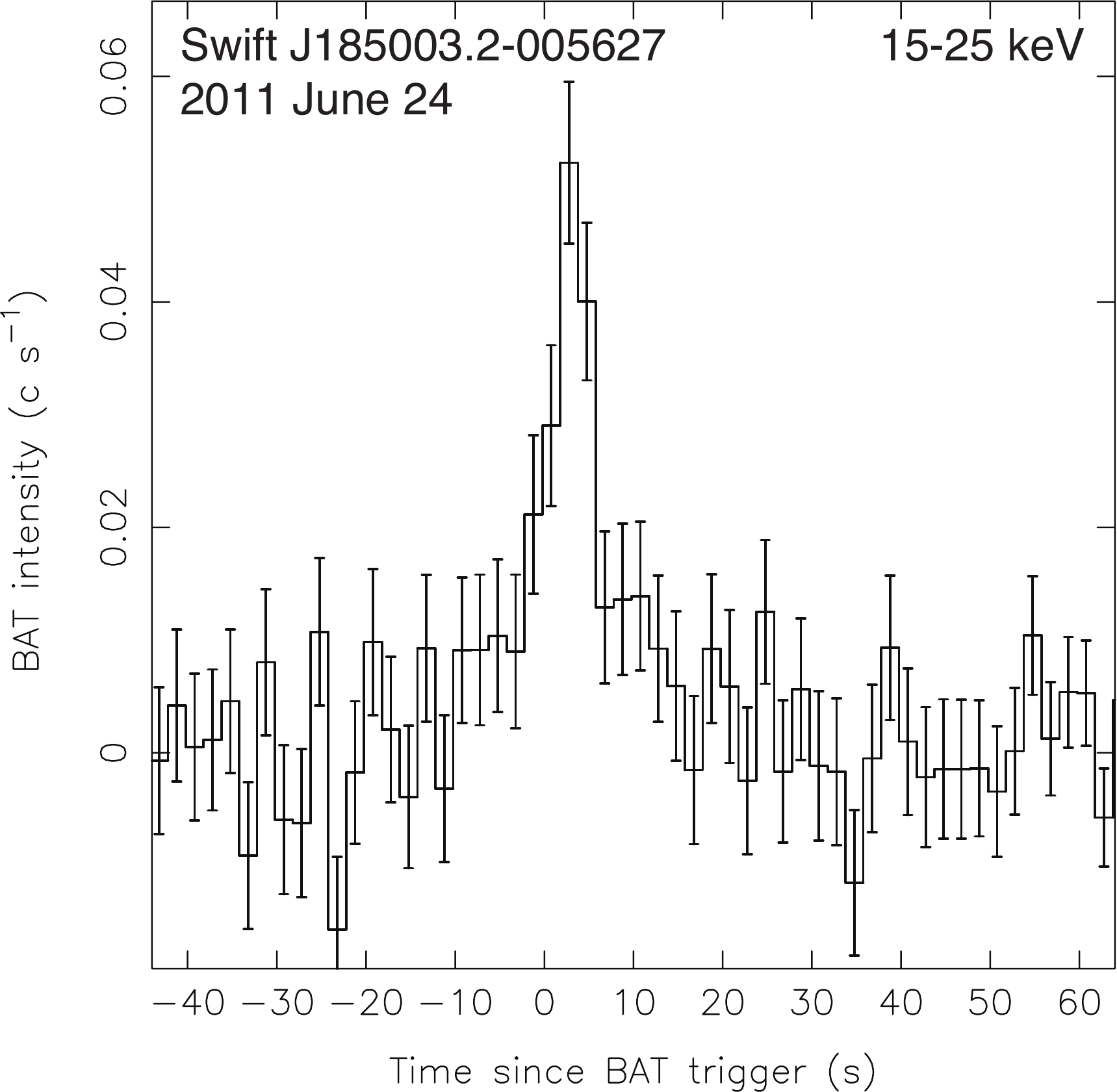}\hspace{0.1cm}
\includegraphics[width=5.8cm]{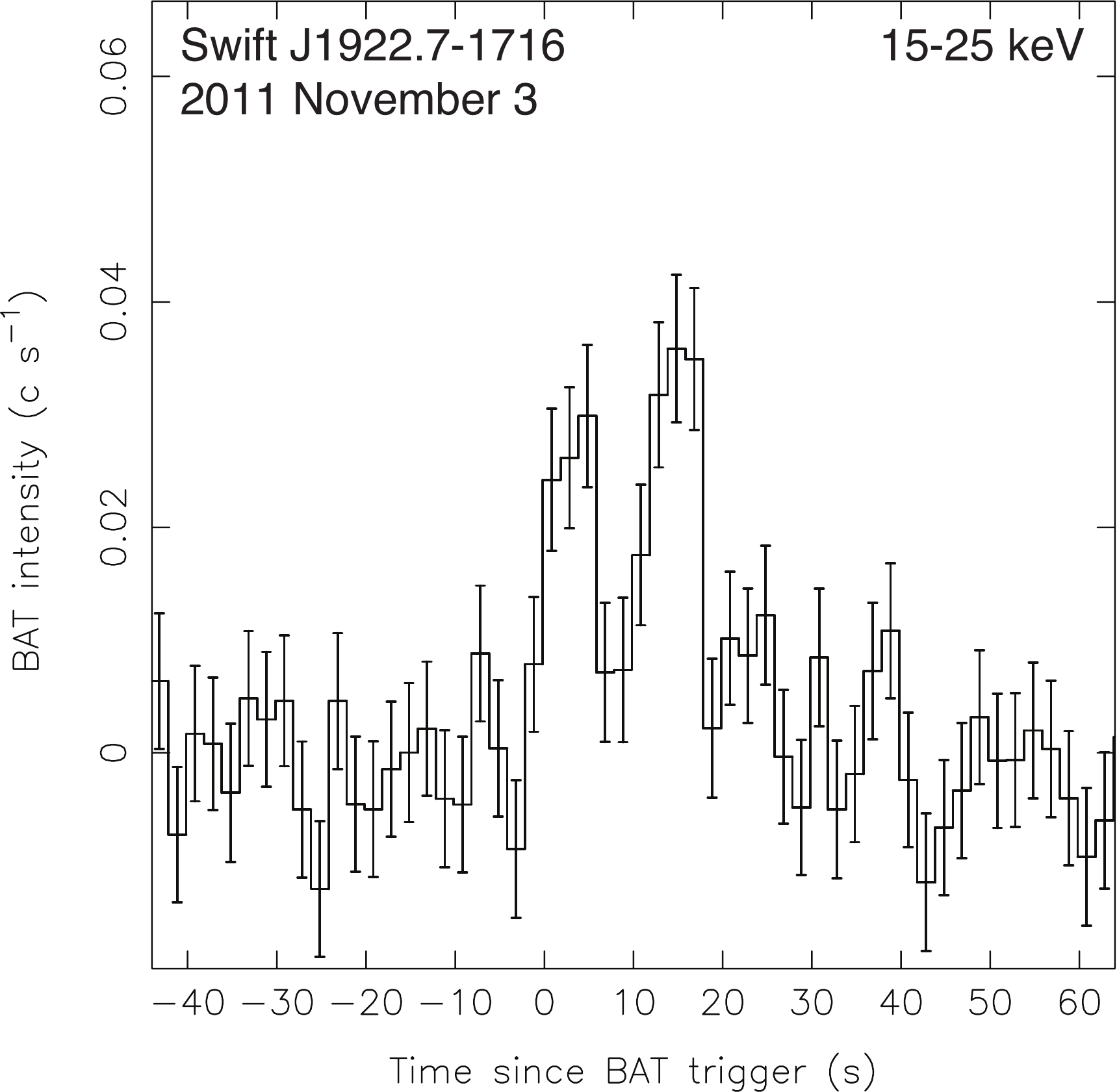}\hspace{0.1cm}
\includegraphics[width=5.8cm]{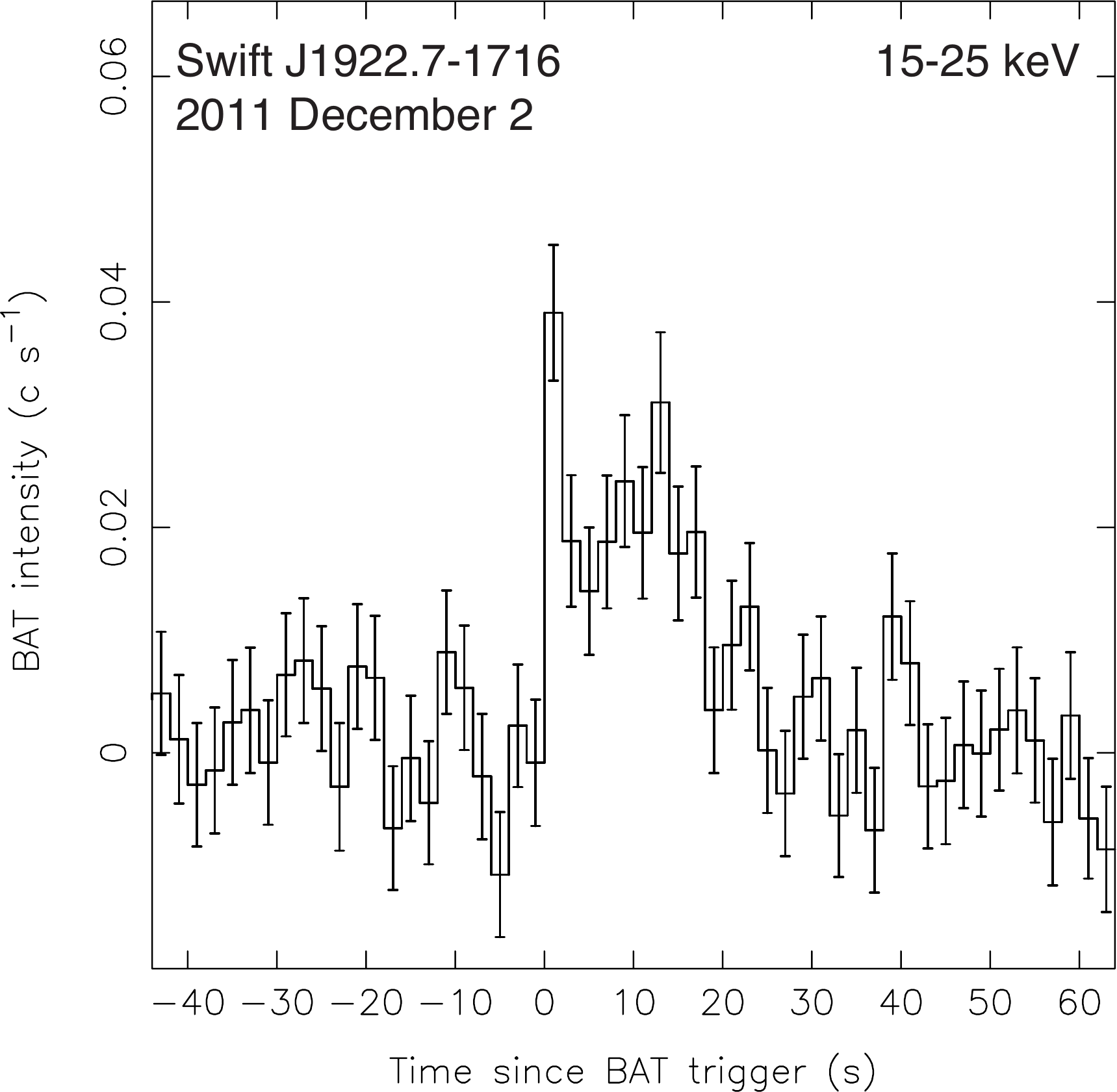}
    \end{center}
\caption[]{{\swift/BAT light curve of trigger 456014 (left), 506913 (middle), and 508855 (right) at 2 s resolution (15--25 keV).}}
 \label{fig:batlc}
\end{figure*}

\begin{figure}
 \begin{center}
\includegraphics[width=8.0cm]{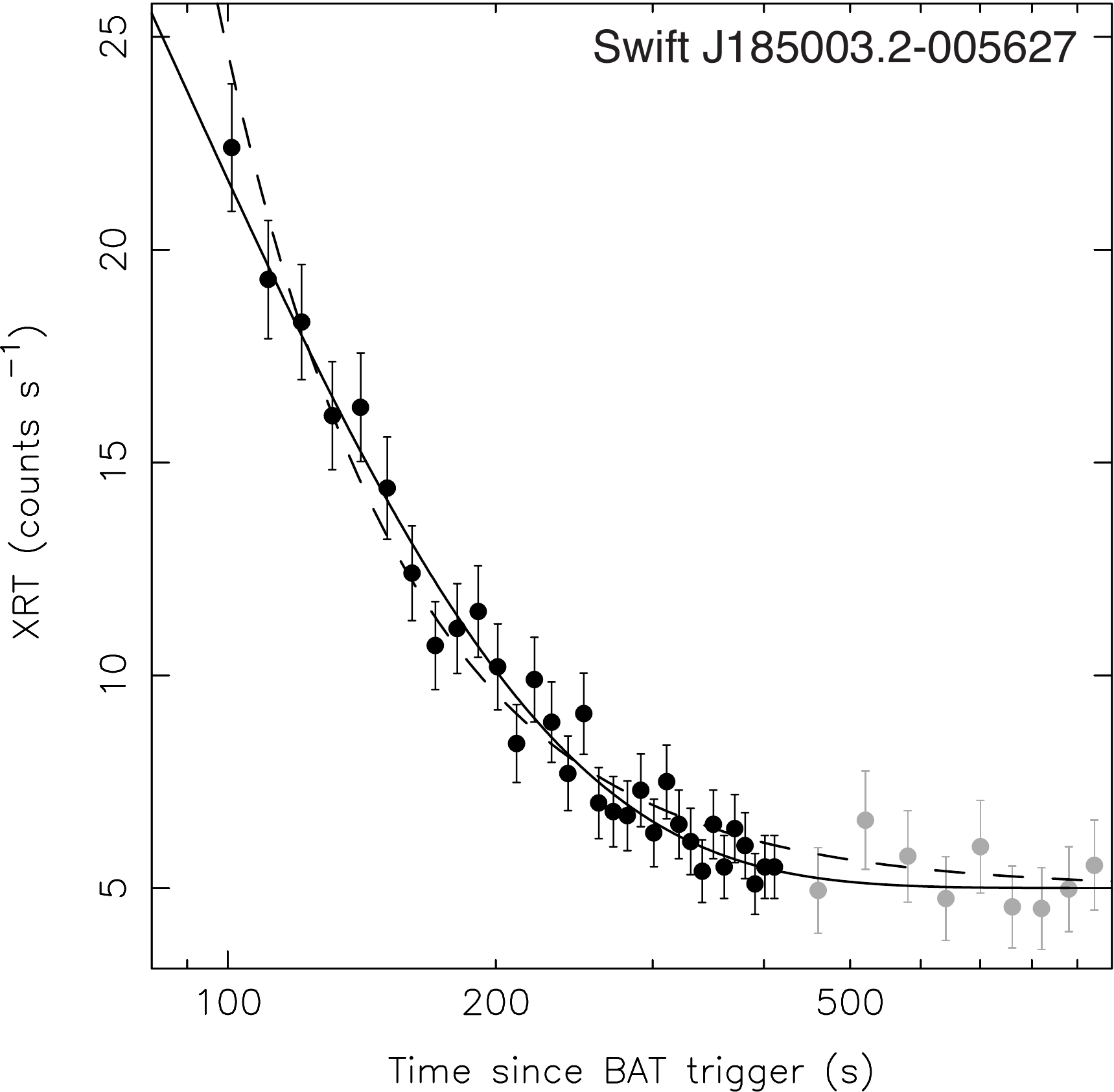}
    \end{center}
\caption[]{{\swift/XRT tail of the X-ray burst of J1850 using a bin time of 10 s. WT data are indicated in black, while PC data are colored gray. The full and dashed curves represent decay fits to an exponential and power-law function, respectively.}}
 \label{fig:1850_decay}
\end{figure}


\subsection{Outburst Properties}\label{subsubsec:new_ob}
\subsubsection{XRT}
Figure~\ref{fig:lc_1850} shows the outburst evolution of J1850 as seen with \swift/XRT. Following the BAT trigger on 2011 June 24, the source was observed almost daily for 27 days till 2011 July 21. In the days following the X-ray burst the intensity remained steady at $\simeq 5~\cnts$ (WT), but the source started fading around 2011 July 5 (11 days after the BAT trigger). It eventually went undetected with the XRT starting on 2011 July 16 (22 days after the BAT trigger). 
Considering the sensitivity of the XRT, a non-detection in $\simeq3.3$~ks of data between 2011 July 16 and 21 (Table~\ref{tab:obs}) suggests that the source luminosity had dropped below a few times $10^{33}~\lum$ (depending on the spectral shape, 0.5--10 keV). This indicates that the outburst had ceased and that the source had returned to quiescence. We further constrain the quiescent emission in Section~\ref{subsubsec:new_q}.

The outburst observed with the XRT had a duration of $\simeq3$~weeks, but the BAT already detected activity from the source on 2011 May 18--26 (see Section~\ref{sec:intro}). If this was part of the same outburst, it had a duration of $\gtrsim8$~weeks. Lack of X-ray data before this epoch does not allow us to constrain the onset of the activity any further. 

To obtain a global characterization of the outburst spectrum and flux we fitted the PC and WT data simultaneously. The hydrogen column density was fixed between the two data sets, whereas the other fit parameters were left to vary freely. The results of our spectral analysis are summarized in Table~\ref{tab:spec}. 

Fitting the spectra with either a single power-law or blackbody model yields reduced chi-square values of $\chi_{\nu}^2 \gtrsim 1.3$ for 673 degrees of freedom (dof). A combined power-law and blackbody model provides a better description of the data and results in $N_{\mathrm{H}} \simeq (1.1\pm 0.1)\times10^{22}~\nh$, $\Gamma \simeq 0.9-1.6$, and $kT_{\mathrm{bb}} \simeq 0.7$~keV ($\chi_{\nu}^2 = 1.1$ for 669 dof; Table~\ref{tab:spec}). The blackbody component contributes $\simeq45\%-65\%$ to the total unabsorbed 0.5--10 keV model flux. These spectral parameters are typical for low-luminosity neutron star LMXBs in the hard state \citep[e.g.,][]{lin2009,linares2009_thesis,armas2011}. The average PC and WT outburst spectra are shown in Figure~\ref{fig:spec}.

The corresponding 0.5--10 keV model flux is $F_{\mathrm{X}} \simeq 1.7 \times10^{-10}~\flux$, when averaged over the PC and WT data. For a distance of $D=3.7$~kpc (see Section~\ref{subsubsec:new_burst}), this translates into a mean outburst luminosity of $L_{\mathrm{X}} \simeq 2.8\times10^{35}~\lum$ (0.5--10 keV). This allows for an estimate of the average bolometric accretion luminosity of $L_{\mathrm{bol}}\simeq 7 \times10^{35}~(D/3.7~\mathrm{kpc})^2~\lum$, which corresponds to $\simeq 0.2$\% of the Eddington limit (Section~\ref{subsec:ana}). 

To investigate whether any spectral evolution occurred along the X-ray outburst, we compared the ratio of counts in the 2--10 keV and 0.5--2 keV energy bands (Figure~\ref{fig:hardness}). This suggests that the spectrum softens (i.e., the hardness ratio becomes smaller) as the intensity decreases. This behavior has been seen for several black hole and neutron star LMXBs transitioning from the hard state to quiescence \citep[][and references therein]{armas2011}.

\subsubsection{UVOT}
Investigation of the UVOT images does not reveal a candidate optical/UV counterpart for J1850 \citep[see also][]{beardmore2011_gcn}. This may not be surprising given the relatively high extinction in the direction of the source. Using the relation $N_{\mathrm{H}}/A_{\mathrm{V}} = (2.21 \pm 0.03)\times 10^{21}~\mathrm{atoms~cm}^{-2}$~mag$^{-1}$, the hydrogen column density inferred from spectral fitting ($N_{\mathrm{H}} \simeq 1.1 \times10^{22}~\nh$), suggests an extinction of $A_{\mathrm{V}} \simeq 5.0$~mag in the $V$-band \citep[][]{guver2009}. Due to the proximity of a nearby bright source, we cannot obtain reliable upper limits for J1850 from the UVOT data. 

We note that near-infrared (nIR) observations revealed a possible counterpart with $K\simeq14.8$~mag within the $1.7''$-XRT positional uncertainty, although the probability of a chance detection is considerable in the crowded source region \citep[][]{im2011}. A weak optical source with $I\simeq21$~mag was detected at a position consistent with the candidate nIR counterpart \citep[][]{gorosabel2011}.

\begin{table}
\caption{UVOT Observations and Results for J1922.}
\begin{center}
\begin{threeparttable}
\begin{tabular}{l l l l l}
\toprule
Obs ID & Date & Band & Exp. & Magnitude \\
 &  &  & (ks) &  \\
\midrule
35174001 & 2005 Jul 8 & $um2$ & 1.7 & $16.13\pm0.03$ \\ 
 &  & $u$ & 0.3 & $16.12\pm0.03$ \\ 
 &  & $v$ & 0.5 & $17.11\pm0.06$ \\
 &  & $uw1$ & 1.1 & $16.05\pm0.03$ \\
 &  & $uw2$ & 2.3 &$16.05\pm0.03$  \\    
35174003 & 2005 Oct 1 & $b$ &  & $17.31\pm0.04$ \\ 
 &  & $um2$ & 0.4 & $16.08\pm0.03$ \\ 
 &  & $u$ & 2.7 & $16.17\pm0.03$ \\ 
 &  & $v$ & 0.9 & $16.98\pm0.05$ \\
 &  & $uw1$ & 1.8 & $16.03\pm0.03$ \\
 &  & $uw2$ & 3.6 & $16.04\pm0.02$ \\   
35471001 & 2006 Mar 14 & $uw2$ & 1.5 & $17.62\pm0.03$ \\ 
35471002 & 2006 Jun 4 & $v$ & 0.3 &$17.45\pm0.10$ \\ 
35471003 & 2006 Jun 18 & $v$ & 1.0 & $17.42\pm0.05$ \\ 
35471004 & 2006 Oct 30 & $v$ & 5.1 & $>20.69$ \\ 
35471005 & 2006 Nov 3 & $v$ & 1.0 & $>19.74$ \\ 
35471006 & 2006 Nov 10 & $v$ & 4.2 & $>20.42$ \\ 
35471007 & 2006 Nov 14 & $v$ & 3.8 & $>20.24$ \\ 
35471008 & 2011 Aug 13 & $um2$ & 1.0 & $16.52\pm0.04$ \\ 
35471009 & 2011 Aug 30 & $uw1$ & 1.2 & $16.34\pm0.03$ \\ 
35471010 & 2011 Aug 31 & $u$ & 4.2 & $16.45\pm0.02$ \\ 
506913000 & 2011 Nov 3 & $b$ & 0.2 & $17.13\pm0.06$ \\ 
 &  & $um2$ & 0.2 & $17.80\pm0.06$ \\ 
 &  & $u$ & 0.2 & $16.42\pm0.05$ \\ 
 &  & $v$ & 0.2 & $17.40\pm0.14$ \\
 &  & $uw1$ & 0.2 & $17.64\pm0.05$ \\
 &  & $uw2$ & 0.2 & $16.03\pm0.05$ \\
 &  & $wh$ & 0.4 & $16.64\pm0.04$ \\    
35471011 & 2011 Nov 4 & $b$ & 0.1 & $17.08\pm0.09$ \\ 
 &  & $um2$ & 0.2 & $15.98\pm0.06$ \\ 
 &  & $u$ & 0.1 & $17.02\pm0.06$ \\ 
 &  & $v$ & 0.1 & $17.54\pm0.28$ \\
 &  & $uw1$ & 0.1 & $15.97\pm0.06$ \\
 &  & $uw2$ & 0.3 & $15.85\pm0.04$ \\
35471012 & 2011 Nov 5 & $b$ & 0.1  & $17.31\pm0.11$ \\ 
 &  & $um2$ & 0.2 & $18.15\pm0.07$ \\ 
 &  & $u$ & 0.1 & $16.44\pm0.08$ \\ 
 &  & $v$ & 0.1 & $17.71\pm0.30$ \\
 &  & $uw1$ & 0.2 & $16.26\pm0.07$ \\
 &  & $uw2$ & 0.3 & $16.25\pm0.05$ \\
35471013 & 2012 Mar 9 & $u$ & 0.4 & $17.12\pm0.07$ \\ 
 &  & $uw1$ & 0.3 & $17.11\pm0.09$ \\
35471014 & 2012 Mar 14 & $u$ & 1.0 & $17.08\pm0.04$ \\ 
35471015 & 2012 May 25 & $u$ & 1.2 & $>20.65$ \\ 
35471016 & 2012 Jun 9 & $uw1$ & 1.9 & $>20.94$\\ 
35471017 & 2012 Jun 11 & $uw2$ & 2.0 & $>21.21$ \\ 
35471018 & 2012 Jun 13 & $uw1$ & 0.9 & $>20.64$ \\ 
35471019 & 2012 Jun 15 & $uw2$ & 0.7 & $>20.62$ \\ 
35471020 & 2012 Jun 17 & $uw1$ & 0.3 & $>19.84$ \\ 
35471021 & 2012 Jun 21 & $uw1$ & 2.3 & $>21.14$ \\ 
\bottomrule
\end{tabular}
\label{tab:uvot}
\begin{tablenotes}
\item[]Note. -- Quoted errors on the observed magnitudes correspond to a 1-$\sigma$ confidence level. In case the of a non-detection, a 3 $\sigma$ upper limit is given. 
\end{tablenotes}
\end{threeparttable}
\end{center}
\end{table}


\subsection{X-Ray Burst Properties}\label{subsubsec:new_burst}
\subsubsection{BAT Burst Peak}
Figure~\ref{fig:batlc} displays the 15--25 keV BAT light curve of trigger 456014 at 2-s resolution. The light curve looks like a single peak centered at $t\simeq2$~s after the trigger. The source is visible for $\simeq25$~s from $\simeq10$~s prior to the BAT trigger to $\simeq15$~s thereafter. Slewing of the spacecraft started at $t\simeq35$~s, when the source had already faded into the background. We used an interval of 20 s centered around the peak (covering $t=-5$ till $15$~s) to extract the average BAT spectrum. 

The BAT spectrum is soft, with no source photons detected above $\simeq35$~keV. It fits to a blackbody with $kT_{\mathrm{bb}}\simeq2.3$~keV and $R_{\mathrm{bb}}\simeq5.6$~km, resulting in $\chi_{\nu}^2 = 1.0$ for 7 dof (assuming $D=3.7$~kpc and fixing $N_{\mathrm{H}} = 1.1\times10^{22}~\nh$). Extrapolating the model fit to the 0.01--100 keV energy range yields an estimate of the bolometric flux of $F_{\mathrm{bol}} \simeq 7.5 \times10^{-8}~\flux$. For a duration of $\simeq20$~s, we estimate a (bolometric) fluence of $f_{\mathrm{BAT}} \simeq 1.5 \times10^{-6}~\fluence$ (Table~\ref{tab:burst_spec}).

While the average count rate over the 20-s interval is $2.0\times10^{-2}~\cnts$, the peak of the BAT data is a factor $\simeq3$ higher. We therefore estimate a bolometric peak flux of $F_{\mathrm{bol,peak}} \simeq 2.3 \times10^{-7}~\flux$. Assuming that the peak was equal to the empirical Eddington limit inferred from PRE bursts (Section~\ref{subsec:ana}), places the source at a distance of $D=3.7$~kpc. 
The data do not reveal signatures of PRE, which implies that the burst may have been sub-Eddington. This distance estimate should therefore be regarded as an upper limit. 

\subsubsection{XRT Burst Tail}
Observations with the narrow-field instruments commenced $\simeq100$~s after the BAT trigger. The inset of Figure~\ref{fig:lc_1850} displays the XRT data, which shows a clear decay in count rate from $\simeq20~\cnts$ at the start of the observation, leveling off to $\simeq5~\cnts$ about 300 s later. This suggests that the total duration of the X-ray burst was $\simeq400$~s ($\simeq7$~minutes). 

The statistics of the XRT data does not allow for a detailed time-resolved spectroscopic analysis. To investigate whether the data exhibit the characteristic spectral softening seen in the tails of X-ray bursts, we extracted two separate spectra for $t=101-201$ and $t=202-430$~s after the burst trigger (all WT data). 

We use an interval of $\simeq500$~s of PC data, obtained between $t=525$ and $1025$~s, to subtract the underlying accretion emission. Fitting this persistent spectrum to an absorbed power-law model yields an unabsorbed 0.5--10 keV flux of $F_{\mathrm{X}}\simeq4.8\times10^{-10}~\flux$. This suggests that J1850 was accreting at $\simeq0.5\%$ of the Eddington rate when the X-ray burst occurred.

The XRT spectra along the burst tail are best described by an absorbed blackbody model. We fix $N_{\mathrm{H}} = 1.1 \times10^{22}~\nh$ (Section~\ref{subsubsec:new_ob}), and find that there is evidence of cooling along the $\simeq300$~s decay tail from $kT_{\mathrm{bb}} \simeq 0.81\pm0.03$~keV to $kT_{\mathrm{bb}} \simeq 0.73\pm0.03$~keV, with corresponding blackbody emitting radii of $R_{\mathrm{bb}}\simeq4-5$~km (Table~\ref{tab:burst_spec}). The obtained spectral parameters are typical of X-ray burst tails and provide further support that the BAT was triggered by a thermonuclear event. 

To investigate the shape of the decay, we fitted the XRT count rate light curve (binned by 10~s) to a power-law and an exponential function (Figure~\ref{fig:1850_decay}). For both decay functions, we include a constant offset to represent the underlying persistent X-ray emission. The power-law fit yields a decay index of $\alpha = -2.1\pm0.1$ and a normalization of $\simeq 3.1\times10^{5}~\cnts$ ($\chi_{\nu}^2 = 0.7$ for 30 dof). For the exponential function, we obtain a decay time of $\tau = 84.7\pm4.9$~s and a normalization of $\simeq54~\cnts$ ($\chi_{\nu}^2 = 0.5$ for 30 dof). 

We apply a count rate to flux conversion factor inferred from fitting the average XRT burst spectrum and integrate both decay fits between $t=20$~s (the time when the BAT intensity had returned to the background level) till $t=400$~s (when the XRT intensity had leveled off to its persistent value). This yields an estimate of the fluence along the XRT burst tail of $f_{\mathrm{XRT}} \simeq (3.8-8.5)\times10^{-7}~\fluence$. Adding this to the value obtained for the BAT data suggests a total fluence of $f_{\mathrm{tot}} \simeq (1.9-2.4)\times10^{-6}~\fluence$ for this event (which is corrected for the underlying persistent emission). For a distance of $D=3.7$~kpc, the corresponding radiated energy is $E_b \simeq (3.1-3.9)\times10^{39}$~erg. 

We searched the light curves of all individual XRT observations for occurrences of X-ray bursts. Apart from the 2011 June 24 event, no other bursts were found.


\subsection{Quiescent Luminosity}\label{subsubsec:new_q}
J1850 is within the field of view (FOV) of three archival \xmm\ observations obtained in 2003 (Table~\ref{tab:obs}). The source is not detected in any of these observations. We derive a 95\% confidence upper limit on the count rate of $\lesssim 1\times10^{-2}~\cnts$ for the PN data, and $\lesssim 1\times10^{-3}~\cnts$ for the MOS cameras, after applying the prescription for small numbers of counts given by \citet{gehrels1986}. Using \textsc{pimms} (ver. 4.5), we estimate the corresponding upper limit on the quiescent luminosity. 

Since the quiescent spectral shape is unknown, we explored both a power-law spectral model with $\Gamma=1-3$ and a blackbody of $kT_{\mathrm{bb}}=0.2-0.3$~keV (with $N_{\mathrm{H}}=1.1\times10^{22}~\nh$), which are typical values found for the quiescent spectra of neutron star LMXBs \citep[e.g.,][]{asai1998,rutledge1999,wijnands2005,degenaar2012_amxp}. This results in an estimated upper limit on the 0.5--10 keV quiescent luminosity of $L_q \lesssim(0.5-2.8)\times10^{32}~(D/3.7~\mathrm{kpc})^2~\lum$.


\begin{figure*}
 \begin{center}
\includegraphics[width=12.0cm]{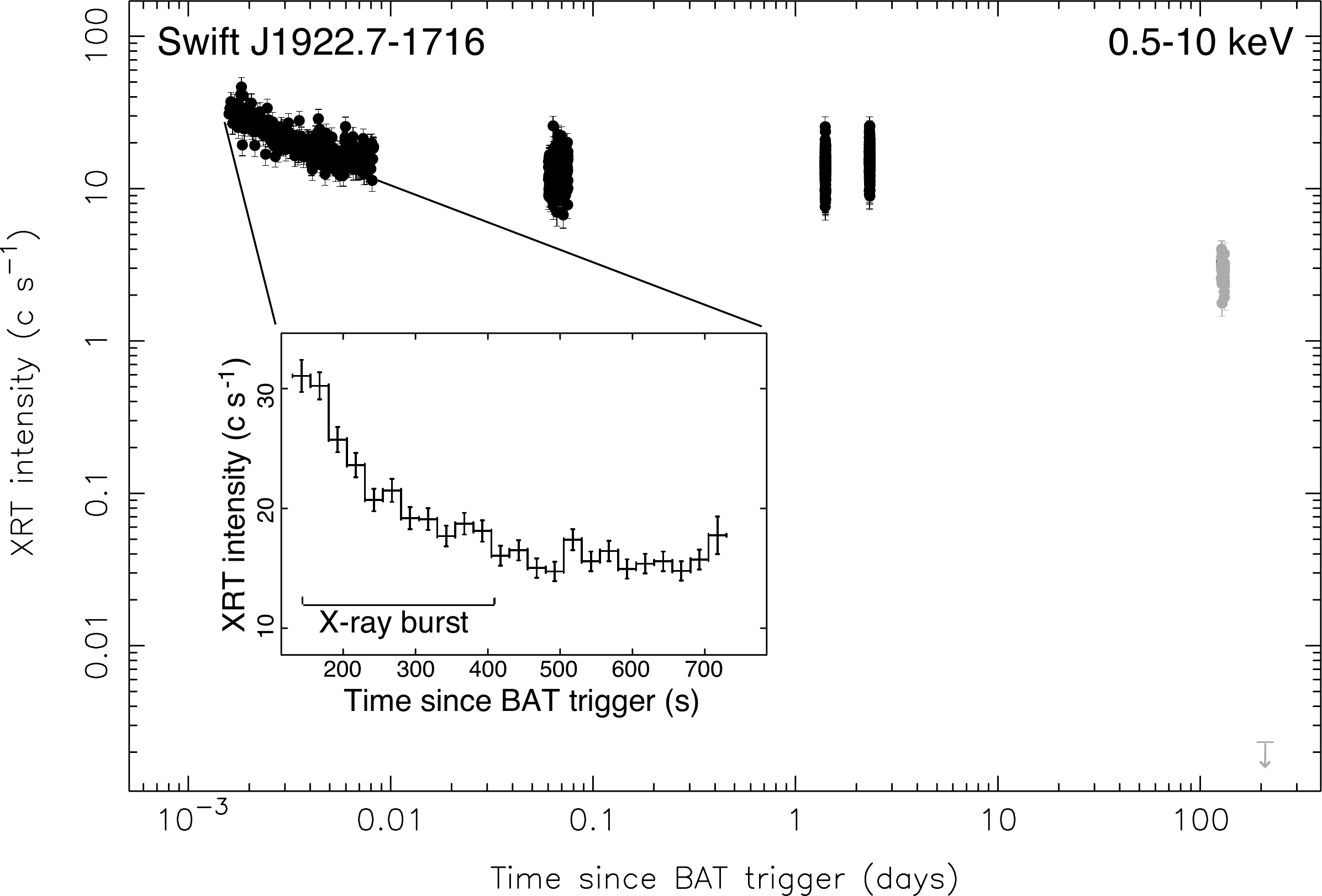}
    \end{center}
\caption[]{{\swift/XRT 0.5--10 keV count rate light curves of J1922 following the BAT trigger (\# 506913) that occurred on 2011 November 3. Both WT (black) and PC (gray) data are used. The main image is a log-log plot showing the X-ray burst and subsequent outburst evolution, using 30 counts per bin. The inset displays the X-ray burst light curve at 25 s resolution on a linear scale. }}
 \label{fig:lc_1922}
\end{figure*}

\begin{figure}
 \begin{center}
\includegraphics[width=8.0cm]{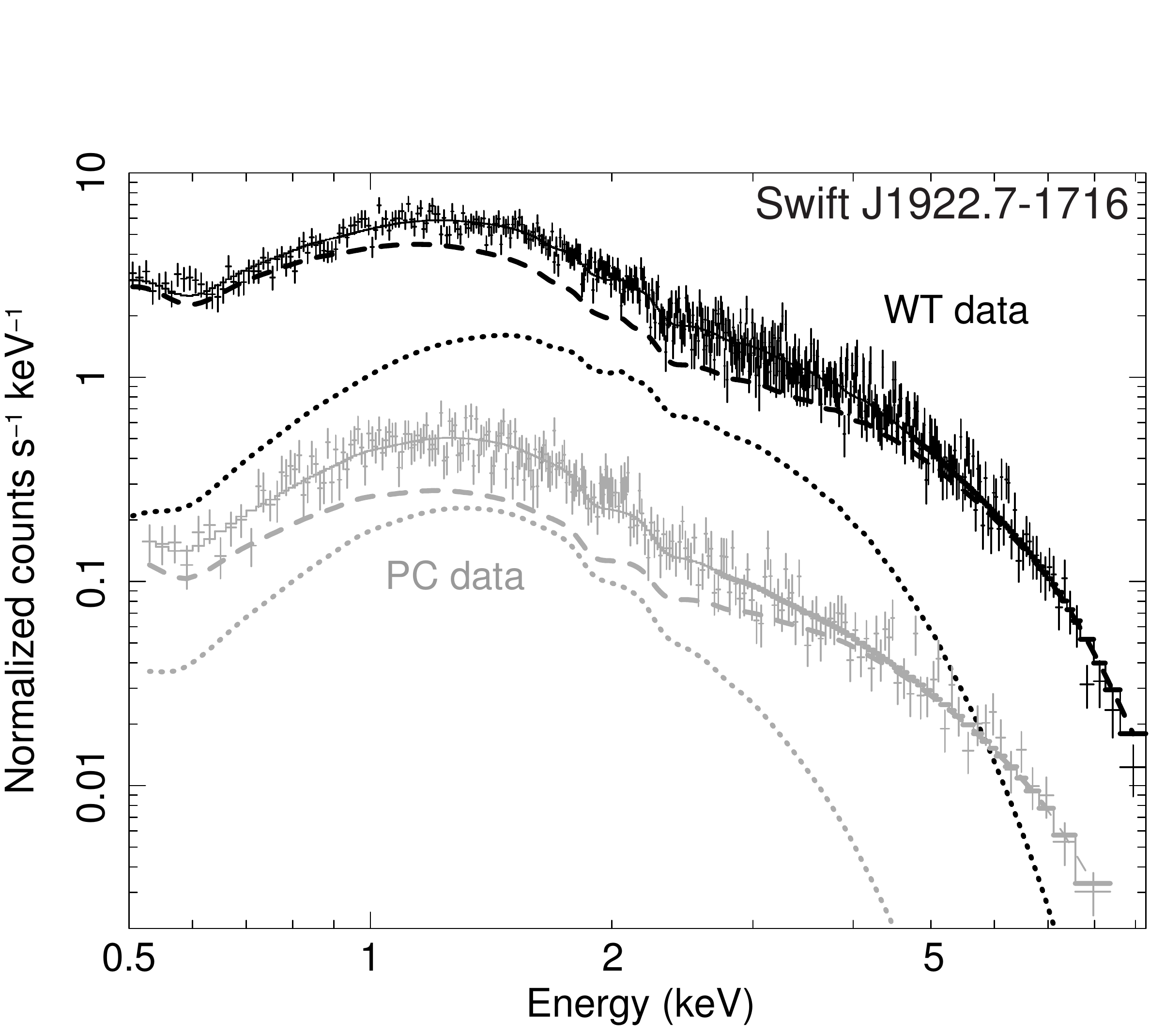}
    \end{center}
\caption[]{{\swift/XRT spectra of the 2011--2012 outburst of J1922, showing both WT (black) and PC (gray) data. The solid lines represent the best fit to a combined power-law (dashed line) and blackbody (dotted curve) model. 
}}
 \label{fig:spec_1922}
\end{figure}

\begin{figure*}
 \begin{center}
\includegraphics[width=8.0cm]{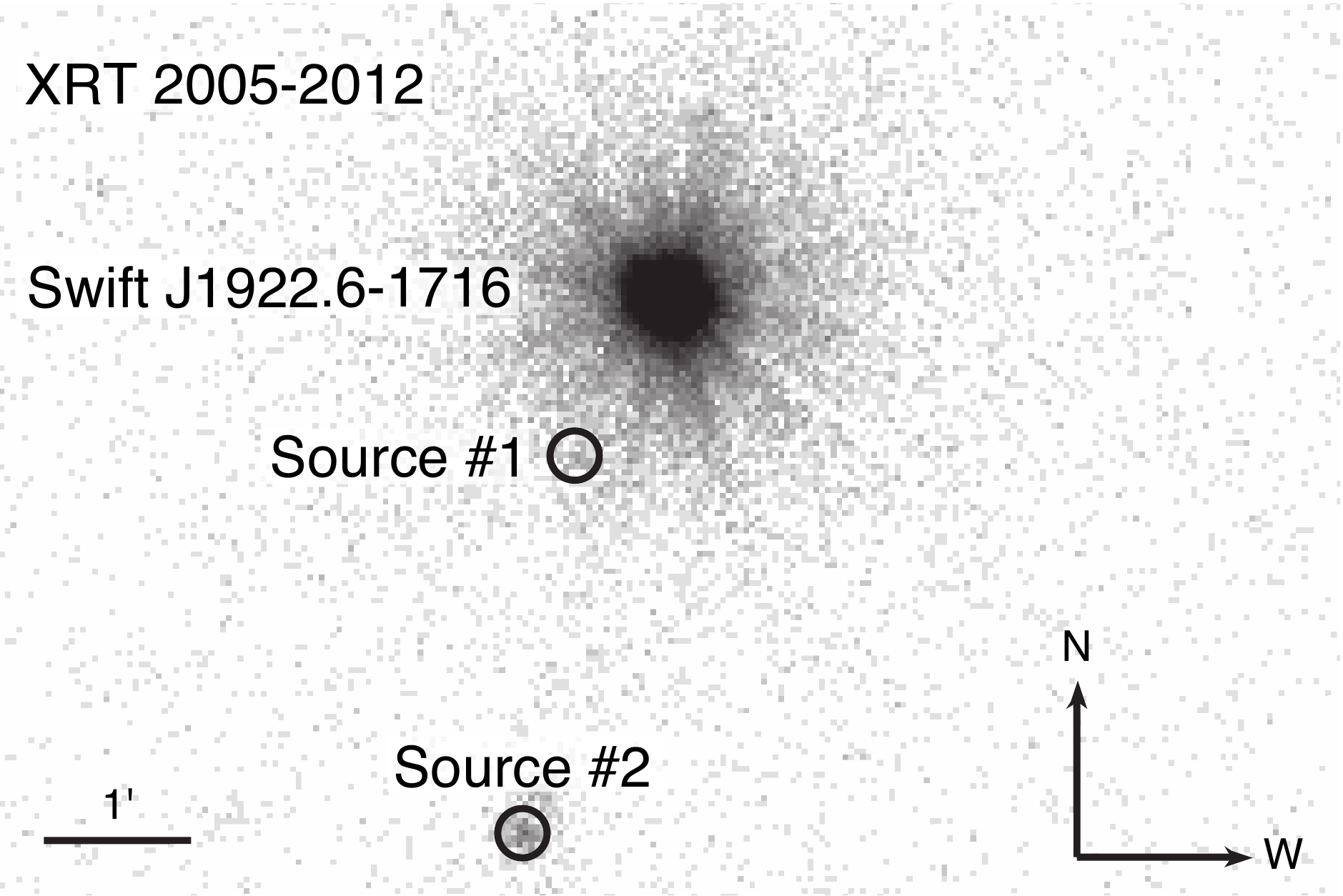}\hspace{0.1cm}
\includegraphics[width=8.0cm]{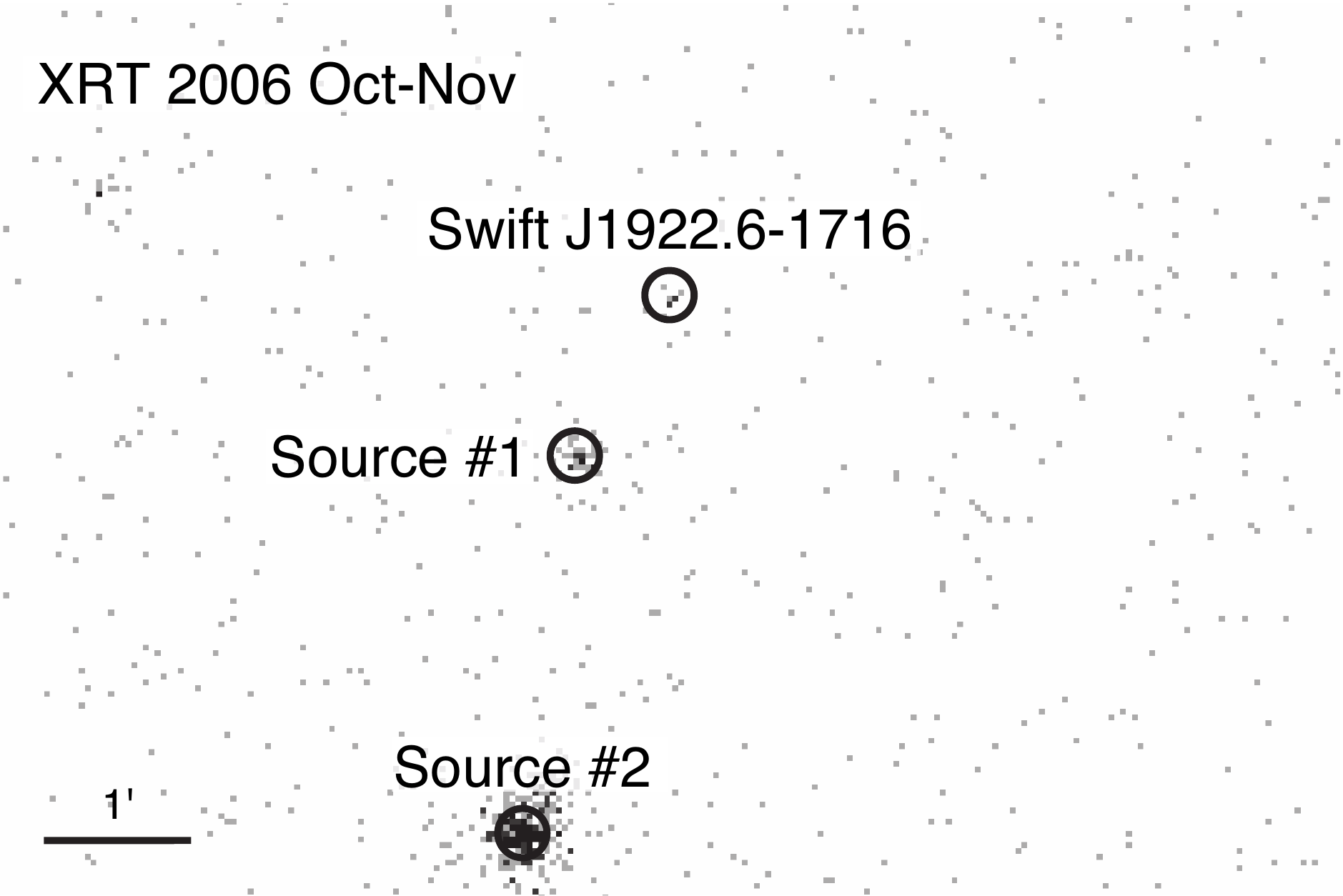}\vspace{0.1cm}
\includegraphics[width=8.0cm]{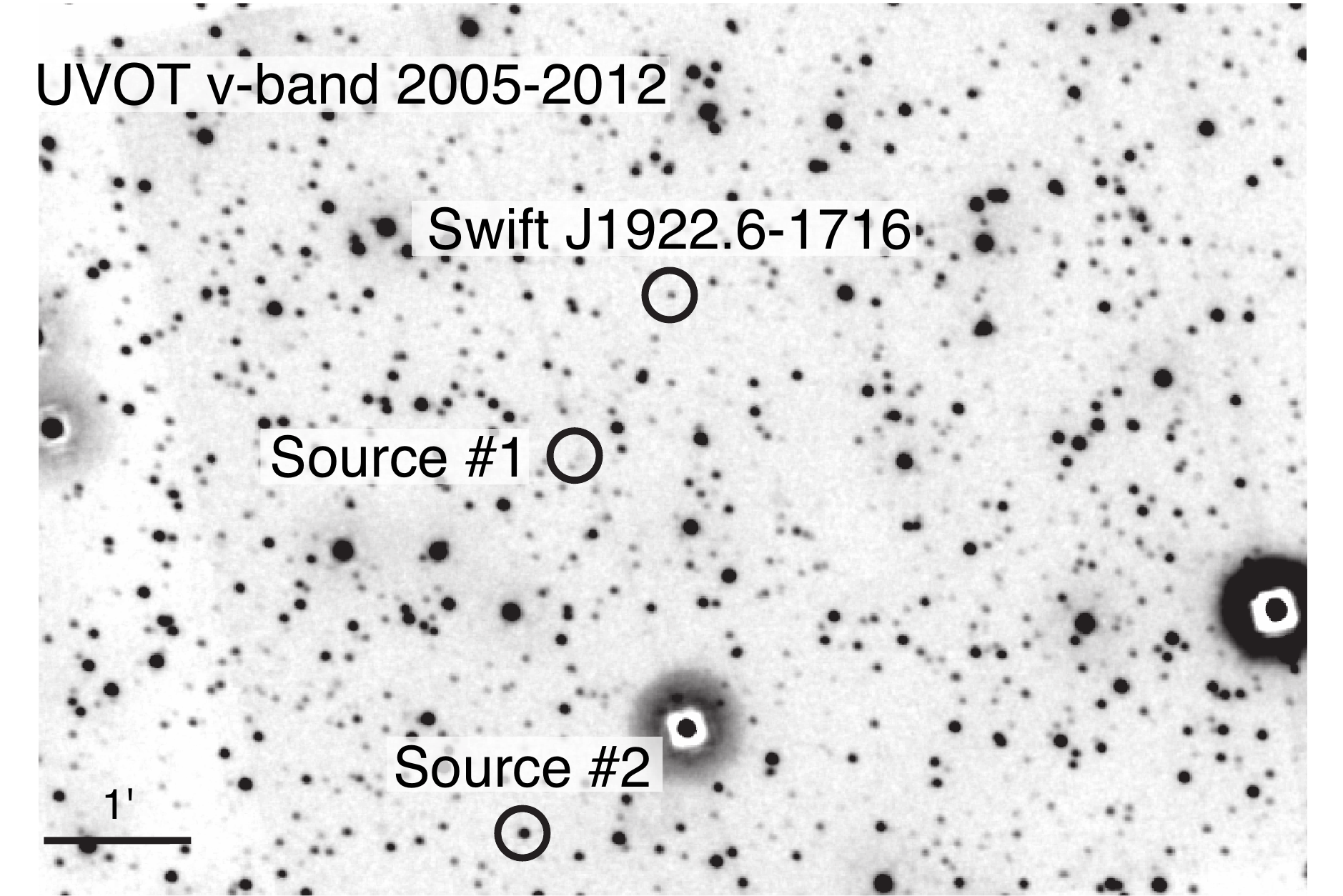}\hspace{0.1cm}
\includegraphics[width=8.0cm]{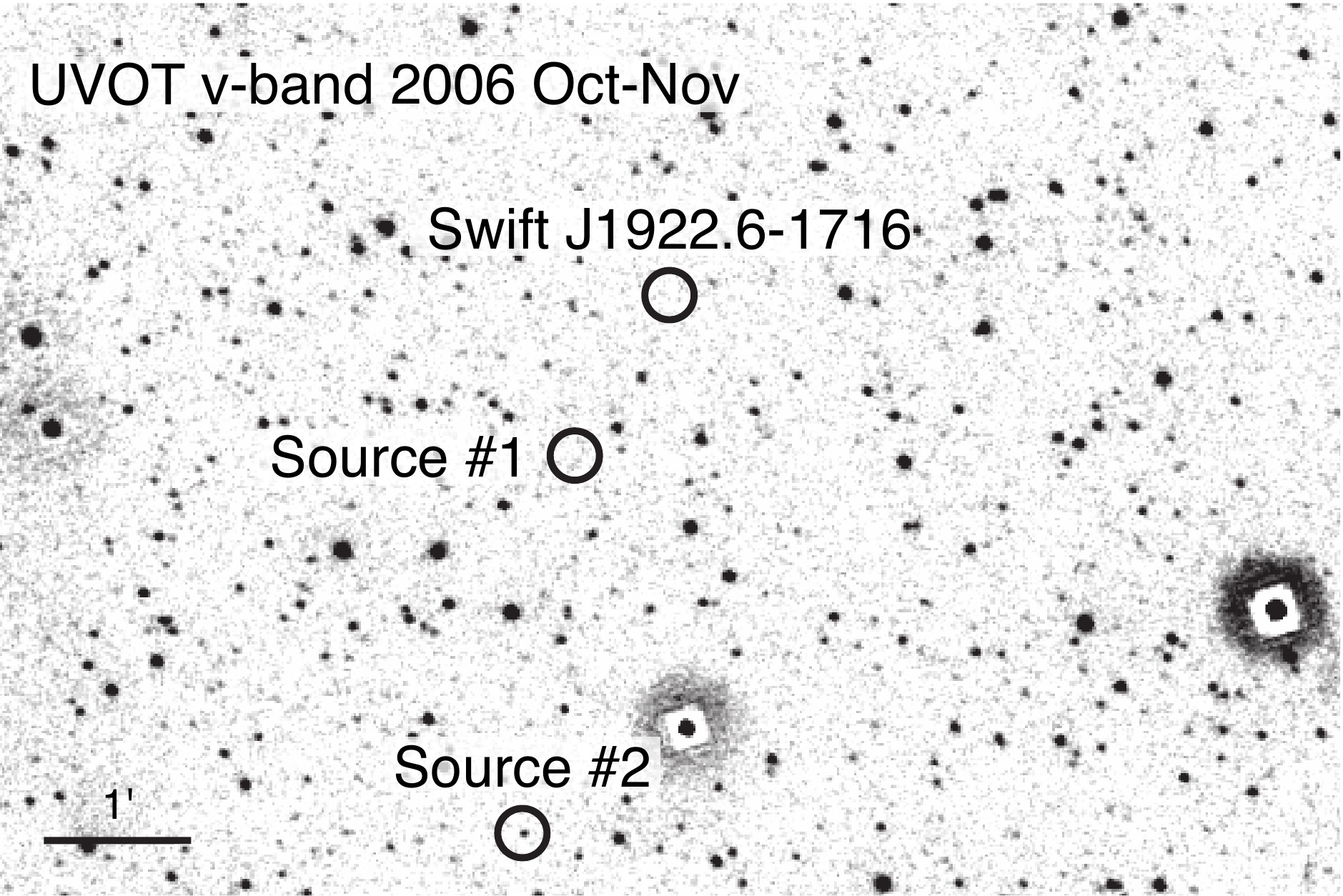}
    \end{center}
\caption[]{{Images of the region around J1922. The location of J1922 and two nearby (uncataloged) X-ray sources are indicated by circles (see the Appendix for details on these serendipitous sources). Top left: accumulated image of all available \swift/XRT data (0.5--10 keV). Top right: \swift/XRT image of 2006 October--November. Bottom left: summed \swift/UVOT $v$-band image. Bottom right: \swift/UVOT $v$-band image of 2006 October--November.}}
 \label{fig:ds9_1922}
\end{figure*}

\section{Results for Swift J1922.7--1716}\label{subsec:source_results}


\subsection{Outburst Properties}\label{subsubsec:source_ob}
\subsubsection{XRT}
\swift\ covered two different outbursts of J1922 in 2005--2006 and 2011--2012. The source was found active in all XRT observations that were carried out between 2005 July 8 and 2006 June 18. However, it is not detected with the XRT in a set of pointings taken between 2006 October 30 and November 14. This demonstrates that the source had returned to quiescence (Section~\ref{subsubsec:source_q}). 
The XRT data suggest that J1922 was active for at least 345 days in 2005--2006. 

Hard X-ray monitoring data also indicate that the source experienced a long outburst. The BAT light curve (15--150 keV) shows that it appeared active right at the start of the survey \citep[2004 December;][]{tueller2005a,tueller2005b}, until it faded $\simeq20$ months later in 2006 July. Examination of the pre-processed \inte/IBIS-ISGR light curve of J1922 (17--80 keV), which is publicly available via the High-Energy Astrophysics Virtually ENlightened Sky \citep[HEAVENS;][]{lubinski2009,walter2010},\footnote{http://www.isdc.unige.ch/heavens/} suggests a similar outburst length as inferred from the BAT data. This is in agreement with the XRT results.

The average XRT spectrum of the 2005--2006 outburst can be described by a single absorbed power-law model, yielding $N_{\mathrm{H}} \simeq (2.9\pm 0.3)\times10^{21}~\nh$ and $\Gamma = 2.1 \pm 0.1$ ($\chi_{\nu}^2 = 0.9$ for 213 dof). Analysis of the broadband 2005 spectrum required the addition of a soft emission component \citep[][]{falanga2006}. We obtain similar results for the 2011--2012 data (see below). Therefore, we included a blackbody component to the power-law fit for comparison.

The combined power-law and blackbody fit results in $N_{\mathrm{H}} \simeq (1.7\pm 0.1)\times10^{21}~\nh$, $\Gamma \simeq 1.5-1.6$, $kT_{\mathrm{bb}} \simeq 0.4 \pm 0.1$~keV, $R_{\mathrm{bb}} \simeq 6-7$~km (for $D=4.8$~kpc), and $\chi_{\nu}^2 = 0.9$ for 209 dof (Table~\ref{tab:spec}). The blackbody component contributes $\simeq20$\% to the total 0.5--10 keV unabsorbed flux. These results are similar to those obtained for the broadband 2005 spectrum \citep[][]{falanga2006}.

The 0.5--10 keV unabsorbed flux (averaged over the PC and WT data) for the combined model fit is $F_{\mathrm{X}} \simeq 2.5 \times10^{-10}~\flux$. This translates into a mean outburst luminosity of $L_{\mathrm{X}} \simeq 6.9 \times10^{35}~\lum$ for a distance of $D=4.8$~kpc (Section~\ref{subsubsec:source_burst}). Assuming that the bolometric luminosity is a factor $\simeq2.5$ higher than the intensity in the 0.5--10 keV band, we estimate an average accretion luminosity of $L_{\mathrm{bol}}\simeq 1.7\times10^{36}~(D/4.8~\mathrm{kpc})^2~\lum$, or $\simeq 0.5$\% of the Eddington limit (Section~\ref{subsec:ana}). 

\swift/BAT and \maxi\ monitoring observations revealed that J1922 entered a new outburst around 2011 July 17 \citep[][]{kennea2011_bursters}. Between 2011 August 13 and 2012 March 15, J1922 was observed during a number of XRT pointings that all detected it in outburst. However, the source is no longer seen with the XRT on 2012 May 25 and follow-up observations performed between 2012 June 9 and 21 (Table~\ref{tab:obs}). This suggests that the source had returned to quiescence (Figure~\ref{fig:lc_1922} and Section~\ref{subsubsec:source_q}). The 2011--2012 outburst had a duration of $\gtrsim240$~days (0.7 yr). If the \swift/BAT hard transient monitor caught the start of it, the duration is constrained to $\lesssim310$~days (0.9 yr). The pre-processed publicly available IBIS-ISGR data do not cover this second outburst from J1922.

Fitting the average XRT spectrum of the 2011--2012 data with a simple absorbed power-law model yields $N_{\mathrm{H}} \simeq (2.7\pm 0.1)\times10^{21}~\nh$ and $\Gamma \simeq 2.0 \pm 0.1$ ($\chi_{\nu}^2 = 1.2$ for 664 dof). The fit can be improved by adding a blackbody component, which results in $N_{\mathrm{H}} \simeq (1.5\pm 0.1)\times10^{21}~\nh$, $\Gamma \simeq 1.6-1.7$, $kT_{\mathrm{bb}} \simeq 0.5-0.7$~keV, $R_{\mathrm{bb}} \simeq 4-5$~km (for $D=4.8$~kpc), and $\chi_{\nu}^2 = 1.0$ for 660 dof (Table~\ref{tab:spec}). The 2011--2012 spectral data and model fit are shown in Figure~\ref{fig:spec_1922}.

The best fit yields an average 0.5--10 keV luminosity for the 2011--2012 outburst of $L_{\mathrm{X}} \simeq 1.1\times10^{36}~(D/4.8~\mathrm{kpc})^2~\lum$, and an estimated bolometric accretion luminosity of $L_{\mathrm{bol}}\simeq 2.7\times10^{36}~(D/4.8~\mathrm{kpc})^2~\lum$ ($\simeq 1$\% of the Eddington limit). The blackbody component contributes $\simeq20$\% to the total unabsorbed 0.5--10 keV model flux. The spectral properties and intensity of the 2011--2012 outburst are comparable to that observed in 2005--2006 (Table~\ref{tab:spec}). 

Investigation of the ratio of counts in different energy bands (0.5--2 and 2--10 keV) does not reveal any evident spectral evolution along the outburst. It is of note that the dynamical range spanned by the observations of J1922 is narrower than that of J1850, for which the spectrum appeared to be softening with decreasing intensity (Section~\ref{subsubsec:new_ob}).

\subsubsection{UVOT}
Inspection of UVOT data reveals a possible counterpart at the position of J1922 in all filters (Table~\ref{tab:uvot}). This object is not present in Digitized Sky Survey (DSS) images \citep[][]{barthelmy2011}, and is not detected during the observations in which J1922 had faded into X-ray quiescence (2006 October--November and 2012 May--June; see Figures~\ref{fig:ds9_1922} and~\ref{fig:xrt_uvot}). This strongly suggests that the source detected in the UVOT images represents the optical/UV counterpart of J1922. As such, the UVOT data provide a sub-arcsecond localization of the LMXB \citep[][]{barthelmy2011}. 

We have listed the magnitudes and upper limits of all UVOT observations in Table~\ref{tab:uvot}. These magnitudes are not corrected for interstellar extinction. Fits to the 2005--2006 and 2011--2012 XRT spectra yield $N_{\mathrm{H}} \simeq 1.6 \times10^{21}~\nh$, which suggests a visual extinction of $A_{\mathrm{V}} \simeq 0.7$~mag \citep[][]{guver2009}. The upper limits obtained with the UVOT for the quiescent state ($v\gtrsim20$~mag, Table~\ref{tab:uvot}) are consistent with the constraints from ground-based telescopes \citep[$>23.4$~mag in the $r$ and $g$ bands;][]{halpern2011}.


\subsection{X-Ray Burst Properties}\label{subsubsec:source_burst}
\subsubsection{BAT Burst Peak}
Figure~\ref{fig:batlc} displays the 15--25 keV BAT light curve of trigger 506913 at 2-s resolution. The light curve of J1922 starts rising $\simeq2$~s before the BAT trigger and shows two separate peaks at $t\simeq5$ and $\simeq15$~s. The source was visible in the BAT for $\simeq 25$~s and had returned to the background level well before \swift\ started slewing ($\simeq65$~s post-trigger). 

We extracted an average BAT spectrum using data from $t=0-20$~s, and fitted this spectrum between 15 and 35 keV. 
Using an absorbed blackbody model yields $kT_{\mathrm{bb}} \simeq 2.4$~keV and $R_{\mathrm{bb}}\simeq6.7$~km (for $D=4.8$~kpc and using $N_{\mathrm{H}} = 1.6 \times10^{21}~\nh$; Table~\ref{tab:burst_spec}). 
Extrapolating the fit to the 0.01--100 keV energy range results in an estimated average bolometric flux of $F_{\mathrm{bol}}\simeq6.8\times10^{-8}~\flux$, and a fluence over the 20-s BAT interval of $f_{\mathrm{BAT}}\simeq1.4\times10^{-6}~\fluence$. These spectral properties and energetics suggest that the BAT triggered on a thermonuclear event.

The BAT light curve shows a double-peaked structure \citep[see also][]{barthelmy2011}. This may be a signature of a PRE phase, suggesting that the burst reached the Eddington limit. However, the limited statistics prohibit confirmation via spectral analysis. The peak count rate seen by the BAT is a factor of $\simeq2$ higher than the average count rate. We therefore estimate that the bolometric flux peaked at $F_{\mathrm{bol,peak}}\simeq1.4\times10^{-7}~\flux$. If the peak reached the empirical Eddington limit of PRE bursts \citep[$\simeq3.8\times10^{38}~\lum$;][]{kuulkers2003} the source distance would be $D=4.8$~kpc. 

The BAT detected another event from the source region one month later on 2011 December 2 (trigger 508855).\footnote{See http://gcn.gsfc.nasa.gov/notices$\_$s/508855/BA/} The light curve of this trigger has a similar duration, intensity and double-peaked structure as the one that occurred in 2011 November (Figure~\ref{fig:batlc}). We estimate a bolometric peak flux of $F_{\mathrm{bol,peak}}\simeq1.7\times10^{-7}~\flux$ and a fluence of $f_{\mathrm{BAT}}\simeq3.4\times10^{-6}~\fluence$. Given that the spectral properties and energetics of the two events are very similar (Table~\ref{tab:burst_spec}), we suggest that BAT trigger 508855 was also caused by an X-ray burst from J1922. No automated follow-up observations occurred for this trigger.

\subsubsection{XRT Burst Tail}
XRT follow-up observations commenced $\simeq136$~s after the BAT trigger of 2011 November 3. The XRT data were carried out in WT mode and show a clear decay in count rate during the first $\simeq300$~s. The intensity peaks at the start of the observation at $\simeq30~\cnts$, after which it gradually decreases and levels off at $\simeq15~\cnts$. In subsequent observations the count rate of J1922 remained at that level (Figure~\ref{fig:lc_1922}). 

We extracted two separate spectra along the 300 s decay, and used the last $\simeq165$~s of the observation as a background reference (all WT data; Figure~\ref{fig:lc_1922}). The spectra can be adequately fitted with a blackbody model that evolves from $kT_{\mathrm{bb}}\simeq0.80 \pm 0.06$~keV in the first 75~s to $kT_{\mathrm{bb}}\simeq0.59 \pm 0.09$~keV in the subsequent 225 s of the decay (for $N_{\mathrm{H}}=1.6\times10^{21}~\nh$; Table~\ref{tab:burst_spec}). Such temperatures are typical for the cooling tail of an X-ray burst, further supporting the idea that the BAT triggered on a thermonuclear event. Investigation of the post-burst persistent spectrum yields a 0.5--10 keV unabsorbed flux of $F_{\mathrm{X}}\simeq 7.8\times10^{-10}~\flux$. This suggests that J1922 was accreting at $\simeq1.4\%$ of the Eddington rate when the burst occurred.

Based on the XRT data, we estimate a total duration for the X-ray burst of $\simeq400$~s (see Fig~\ref{fig:lc_1922}). The count rate light curve can be described by a power law with a decay index of $\alpha = -2.1 \pm 0.2$ and a normalization of $\simeq4.1\times10^{6}~\cnts$ ($\chi^2_{\nu}=1.7$ for 25 dof), or an exponential with a decay time of $\tau=102 \pm 9$~s and a normalization of $\simeq62~\cnts$ ($\chi^2_{\nu}=1.6$ for 25 dof). For both decay fits we included a constant offset to subtract the accretion emission.

We estimate the fluence in the XRT burst tail by applying a count rate to flux conversion factor inferred from fitting the average XRT burst spectrum and integrating the two different decay fits from $t=20$~s (the time when the BAT intensity had faded to the background) till $t=400$~s since the trigger (the time when the XRT count rate had leveled off). This yields $f_{\mathrm{XRT}}\simeq(5.4-9.7)\times10^{-7}~\fluence$ for the burst tail (which is corrected for the underlying persistent emission). Combined with the BAT, this suggests a total burst fluence of $f_{\mathrm{tot}}\simeq(1.9-2.4)\times10^{-6}~\fluence$. This corresponds to a radiated energy of $E_{b}\simeq (5.4-6.5)\times10^{39}$~erg for $D=4.8$~kpc.

\begin{figure*}
 \begin{center}
\includegraphics[width=8.0cm]{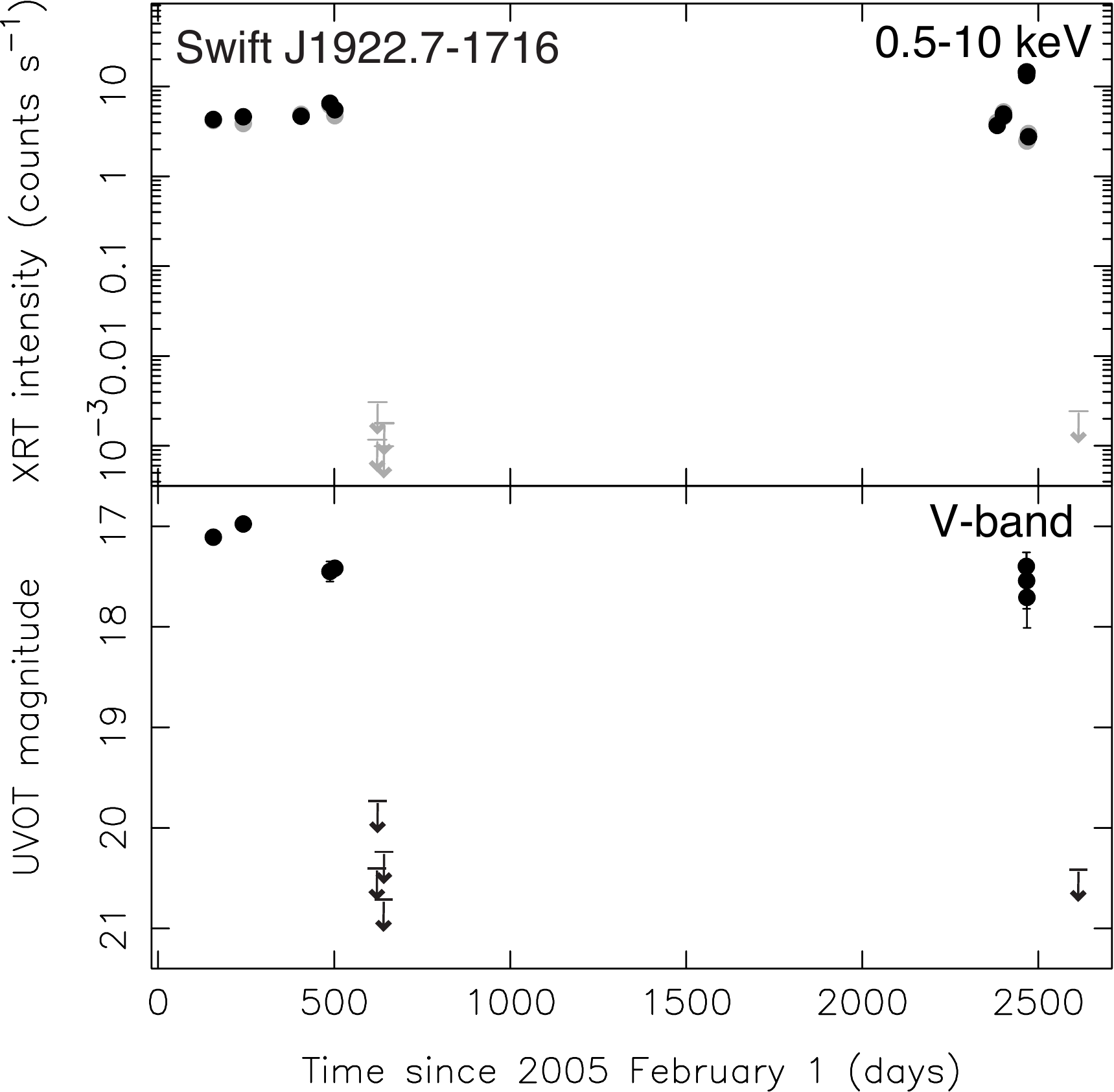}\hspace{0.5cm}
\includegraphics[width=8.0cm]{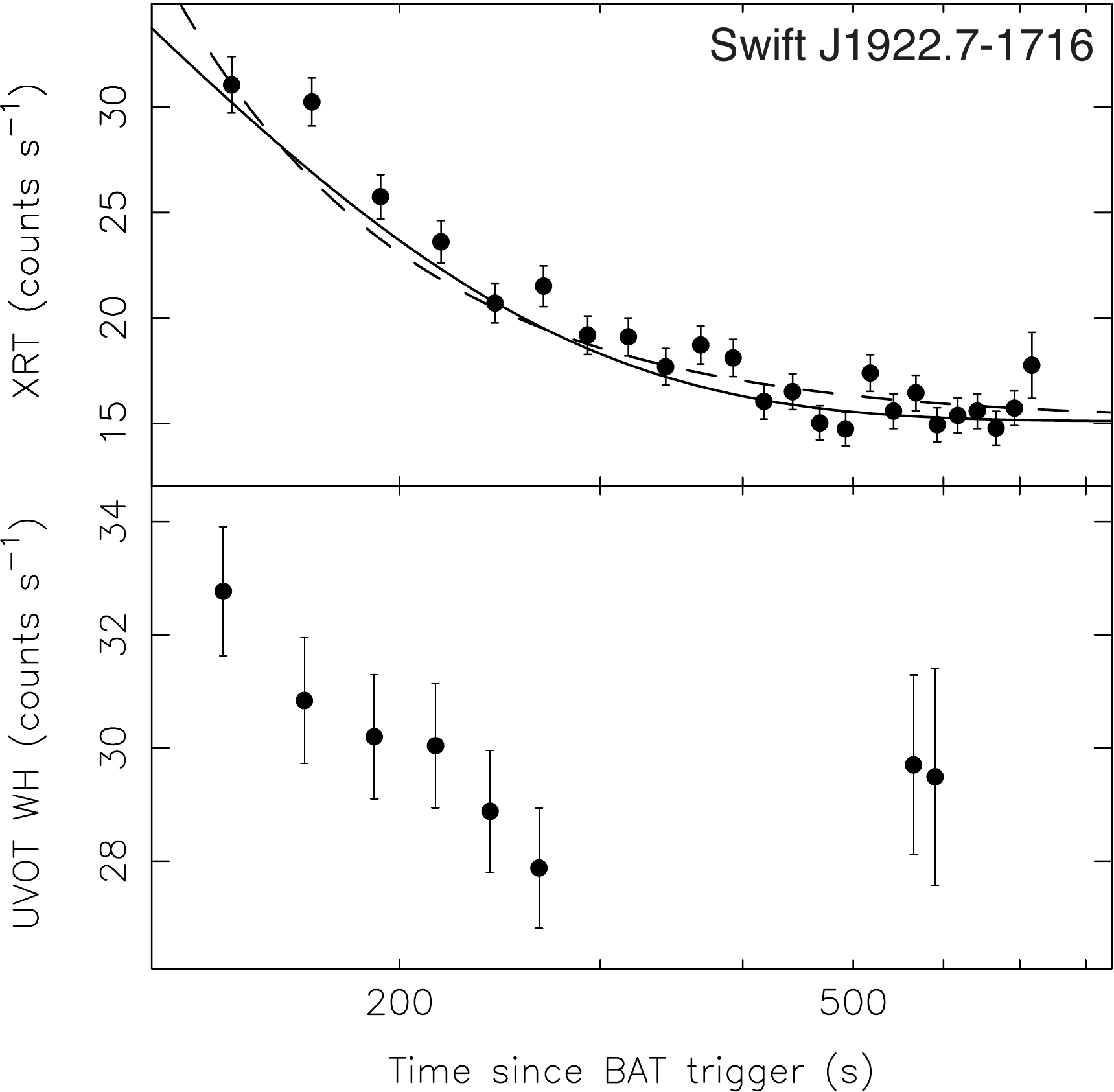}
    \end{center}
\caption[]{{Left: \swift/XRT and UVOT $v$-band data of the two outbursts of J1922. For the XRT, WT data are indicated in black and PC data are colored gray.
Right: \swift/XRT (WT mode) and UVOT $wh$-band data of the follow-up observation of trigger 506913 (2011 November 3), showing the tail of the X-ray burst. The full and dashed curves represent an exponential and power-law decay fit, respectively. 
}}
 \label{fig:xrt_uvot}
\end{figure*}

We note that there are two other X-ray sources located within the XRT FOV of J1922 (Fig~\ref{fig:ds9_1922}). We briefly discuss the properties of these serendipitous X-ray sources in the Appendix. Since the XRT data following BAT trigger 506913 were taken only in WT mode, we cannot formally exclude one of these other X-ray objects as sources of the BAT trigger based on the X-ray data alone. However, we investigated the simultaneous UVOT data of the trigger observation and found that the intensity of J1922 in the UVOT $wh$ band was decreasing simultaneously with the decay seen in the XRT data (Figure~\ref{fig:xrt_uvot}). This provides strong evidence that it was indeed J1922 that triggered the BAT. We did not find any other occurrences of X-ray bursts in the XRT data.

\subsection{Quiescent Luminosity}\label{subsubsec:source_q}
J1922 went undetected during four consecutive XRT observations carried out between 2006 October 30 and November 14 (Obs IDs 35471004--7). We obtain count rate upper limits for these individual exposures of $\lesssim (1-3)\times10^{-3}~\cnts$ (0.5--10 keV). Summing the 4 observations ($\simeq13.8$~ks in total) reveals a small excess of 11 photons at the source position (see Fig~\ref{fig:ds9_1922}), whereas five photons are expected from a set of three randomly placed background regions. 
The inferred count rate is $4.3\times10^{-4}~\cnts$. For $N_{\mathrm{H}}=1.6\times10^{21}~\nh$, a power-law model with $\Gamma=1-3$, or a blackbody with $kT_{\mathrm{bb}}=0.2-0.3$~keV, we estimate a 0.5--10 keV quiescent luminosity of $L_q \simeq (0.4-0.9) \times10^{32}~(D/4.8~\mathrm{kpc})^2~\lum$. 

We note that this quiescent detection occurred within $\simeq 4$ months after the end of a long ($\simeq2$ yr) accretion outburst of J1922. It is possible that the neutron star became considerably heated during this prolonged outburst and did not yet cool down at the time of the \swift/XRT observations (Section~\ref{subsec:q}). In this case, the true quiescent level could be lower than inferred here.

Inspection of an archival 78.6 ks long \suzaku\ observation, performed on 2007 April 10, reveals a possible weak detection at the position of J1922 in the combined XIS image. In the source region, a total of $\simeq450$ photons are detected, whereas $\simeq350$ are expected from the background. We estimate a source count rate of $\simeq 1.3\times10^{-3}~\cnts$. 
Using the range of spectral parameters stated above, we obtain a 0.5--10 keV luminosity of $L_q \simeq (0.6-1.0) \times10^{32}~(D/4.8~\mathrm{kpc})^2~\lum$. This is comparable to our estimate obtained from the \swift/XRT observations.

On 2012 May 25, the source was not detected during a single \swift/XRT observation with an upper limit on the count rate of $\lesssim 2.5\times10^{-3}~\cnts$ at 95\% confidence level \citep[][]{gehrels1986}. This translates into a 0.5--10 keV quiescent luminosity upper limit $L_q \lesssim (2.2-2.5) \times10^{32}~(D/4.8~\mathrm{kpc})^2~\lum$, and strongly suggests that the 2011--2012 outburst had ceased. 

We obtained a series of \swift\ follow-up observations between 2012 June 9 and 21 (Table~\ref{tab:obs}). The source is not detected in the combined XRT image (total exposure time of $\simeq 9.2$~ks when including the observation of 2012 May 25). We infer an upper limit on the count rate of $\lesssim 6.8\times10^{-4}~\cnts$ \citep[95\% confidence;][]{gehrels1986}. Using the range of spectral parameters given above, we obtain a 0.5--10 keV upper limit of $L_q \lesssim (0.4-1.4) \times10^{32}~(D/4.8~\mathrm{kpc})^2~\lum$. 


\section{Burst oscillation search}
Several neutron star LMXBs have shown nearly coherent (millisecond) oscillations during X-ray bursts that measure the spin period of the neutron star \citep[for reviews, see][]{strohmayer06,watts2012}. We searched the 0.5--10 keV \swift/XRT data of the X-ray burst tails for coherent pulsations in the frequency range 1--250 Hz (restricted by the time resolution of the data), using Fourier techniques. We first computed and searched Leahy-normalized \citep{Leahy83} power spectra throughout each burst using 16-s independent segments. This revealed no significant signal.

To estimate upper limits, we used the burst averaged power spectra calculated from a set of Leahy-normalized power spectra with a 1~s window. This takes into account possible frequency drifts (within 1 Hz) and assumes that coherent oscillations are present during the whole burst. We followed \citealt{Vaughan94}, and searched for the largest observed power in the 0.5--10 keV band. Using the Groth-distribution \citep{Groth75}, we obtain upper limits of $\simeq3\%$ and $\simeq2\%$ fractional rms amplitude for J1850 and J1922, respectively ($99.7\%$ confidence level). 
If we assume that the oscillations are only present during the first $\simeq5$~s sampled with the XRT, then the upper limits for both sources are unconstrained ($\gtrsim20$\%).

The low count rates in the \swift/BAT data (Figure~\ref{fig:batlc}) did not allow us to obtain a meaningful upper limits on the occurrence of oscillations during the peak of the X-ray bursts.

\begin{table*}
\caption{Summary of X-Ray Burst Properties}
\begin{center}
\begin{threeparttable}
\begin{tabular}{l c c}
\toprule
Parameter & \new & \source \\
\midrule
Rise time, $t_{\mathrm{rise}}$ (s) & $\simeq 6$ & $\simeq 10$ \\
Exponential decay time, $\tau$ (s) & $\simeq 85$ & $\simeq 102$ \\
Powerlaw decay index, $\alpha$ & $\simeq 2.1$ & $\simeq2.1$ \\
Total duration, $t_{\mathrm{b}}$ (s) & $\simeq 400$ & $\simeq 400$ \\
Total fluence, $f_{\mathrm{tot}}$ ($\fluence$) & $\simeq (1.9-2.4) \times 10^{-6}$ & $\simeq (1.9-2.4) \times 10^{-6}$  \\
Radiated energy, $E_{\mathrm{b}}$ (erg) & $\lesssim (3.1-3.9) \times 10^{39}$ & $\lesssim (5.4-6.5) \times 10^{39}$  \\
Bolometric peak flux, $F_{\mathrm{bol}}$ ($\flux$) & $\simeq 2.3 \times 10^{-7}$ & $\simeq 1.4 \times 10^{-7}$  \\
Distance, $D$ (kpc) & $\lesssim 3.7$ & $\lesssim 4.8$ \\
Post-burst accretion luminosity, $L_{\mathrm{bol}}$ ($\lum$) & $\lesssim 2 \times 10^{36}$ & $\lesssim 5 \times 10^{36}$ \\
Post-burst mass-accretion rate, $\dot{M}$ ($\Msun~\mathrm{yr}^{-1}$) & $\lesssim 2 \times 10^{-10}$ & $\lesssim 4 \times 10^{-10}$ \\
\bottomrule
\end{tabular}
\label{tab:bursts}
\begin{tablenotes}
\item[]Note. The rise time refers to that seen in the BAT. The burst duration is estimated as the time from the BAT trigger till the flux decayed back to the persistent level as observed with the XRT. The quoted peak fluxes are unabsorbed and for the 0.01--100 keV energy range. The quoted accretion luminosity and mass-accretion rate were inferred from fitting $\simeq200-500$~s of post-burst persistent emission. The burst listed for \source\ is the one that triggered the BAT in 2011 November, for which follow-up XRT observations were carried out. We note that the shape of X-ray bursts can look very different in different energy bands, so caution should be taken when comparing rise and decay times seen with different instruments.
\end{tablenotes}
\end{threeparttable}
\end{center}
\end{table*}


\section{Discussion}\label{sec:discuss}
We investigated three \swift/BAT triggers that occurred in 2011. The BAT trigger spectra are soft and can be described by a blackbody model with a temperature of $kT_{\mathrm{bb}}\simeq 2-3$~keV. Rapid follow-up XRT observations were obtained for two of the triggers and revealed a decaying X-ray intensity that levels off to a constant value $\simeq400$~s after the BAT peak. The X-ray tail shows a soft (thermal) emission spectrum that cools during the decay to $kT_{\mathrm{bb}}\simeq 0.6-0.7$~keV. 
Both events have a comparable exponential decay time ($\tau \simeq 85-100$~s), fluence ($f_{\mathrm{total}}\simeq2\times10^{-6}~\fluence$), and radiated energy output ($E_b\simeq 3-7\times10^{39}$~erg; Table~\ref{tab:bursts}). These features are consistent with thermonuclear bursts occurring on accreting neutron stars, and strongly suggest that the BAT triggers were caused by type-I X-ray bursts.

We identified the X-ray transients J1850 and J1922 as the origin of the BAT triggers. This implies that both sources are neutron star LMXBs. J1850 is a previously unknown X-ray source, whereas J1922 was already discovered in 2005 but remained unclassified \citep[][]{falanga2006}. We characterize the outburst and quiescent properties of these two new X-ray bursters. 

The outburst of J1850 seen with \swift's BAT and XRT had a duration of $\simeq8$~weeks, but the onset of the outburst is not well constrained and hence it may have lasted (considerably) longer. The average 0.5--10 keV intensity of $L_{\mathrm{X}}\simeq 3\times10^{35}~(D/3.7~\mathrm{kpc})^2~\lum$ classifies the source as a very-faint X-ray transient \citep[Table~\ref{tab:spec}; cf.][]{wijnands06}. At the time of the X-ray burst, the source was accreting at $\simeq0.5\%$ of the Eddington rate. Investigation of the ratio of XRT counts in different energy bands as the outburst progressed, revealed that the spectrum softens when the intensity decreases. This behavior has been seen in several neutron star and black hole LMXBs transitioning from the hard state to quiescence, although the underlying mechanism is not understood well \citep[see][for a recent example and an overview]{armas2011}.

J1922 exhibited an outburst in 2005--2006 with an observed duration of $\simeq20$~months and an 0.5--10 keV intensity of $L_{\mathrm{X}}\simeq 7\times10^{35}~(D/4.8~\mathrm{kpc})^2~\lum$. The 2011--2012 outburst likely commenced in 2011 mid-July and had ceased by 2012 May, implying a duration of $\simeq8-10$~months. The 0.5--10 keV luminosity of this second outburst was $L_{\mathrm{X}}\simeq1\times10^{36}~(D/4.8~\mathrm{kpc})^2~\lum$. These outburst intensities fall in the very-faint to faint regime \citep[Table~\ref{tab:spec}; cf.][]{wijnands06}. At the time of the BAT trigger of 2011 November, the source was accreting at $\simeq1.4\%$ of the Eddington limit. Using \swift/UVOT data, we have identified a unique UV/optical counterpart to J1922. This provides a sub-arcsecond localization of the source \citep[][]{barthelmy2011}.

\subsection{X-Ray Burst Properties of J1850 and J1922}\label{subsec:long}
The observable properties of X-ray bursts (e.g., the duration, recurrence time and radiated energy) depend on the conditions of the ignition layer, such as the thickness, temperature, and H-abundance. These can drastically change as the mass-accretion rate onto the neutron star varies, such that there exist distinct accretion regimes that give rise to X-ray bursts with different characteristics \citep[][]{fujimoto81,bildsten98}. The overall similarities between the properties of the X-ray bursts of J1850 and J1922 (Table~\ref{tab:bursts}) suggest that these events were ignited under similar conditions. 

In the accretion regime of J1850 and J1922 ($\simeq0.5\%-1\%$ of Eddington), the temperature of the burning layer is expected to be low enough for H to burn unstably. This may in turn trigger He ignition in an H-rich environment, which typically results in $\simeq10-100$~s long bursts \citep[e.g.,][]{fujimoto81,bildsten98,galloway06}. 
The detection of H-emission lines in the optical spectrum of J1922 suggests that the neutron star is accreting H-rich material \citep[][]{wiersema2011,halpern2011}. This implies that the fuel triggering the X-ray burst could have indeed contained H. 
Given the similarities in burst properties, J1850 may therefore be expected to host an H-rich companion star as well. 

The observed burst duration of $t_b\simeq400$~s ($\tau \simeq 85-100$~s) is considerably longer than that typically observed for H-rich X-ray bursts \citep[][]{chenevez2008,linares2012}. The burst profiles may look very different depending on the energy band in which they are observed \citep[][]{chelovekov2006,linares2012}, and most bursts known to date have been detected with instruments that cover higher energies than \swift/XRT \citep[typically $>2$~keV, e.g., \rxte, \inte, \beppo, \fermi;][]{cornelisse2003,chenevez2008,galloway06,linares2012}. However, there are several X-ray bursts observed with \swift/XRT, \chan, or \xmm\ (i.e., in the same energy band as the bursts observed from J1850 and J1922) from neutron star LMXBs accreting at $\simeq1\%$ of Eddington that do show the expected duration of $\simeq10-100$~s \citep[e.g.,][]{boirin2007,trap09,degenaar09_gc,degenaar2012_gc}. 

The uncommon properties of the bursts from J1850 and J1922 are further illustrated by their energetics. The estimated radiated energies of $E_b\lesssim(3-7)\times10^{39}$~erg are higher than that typically found for normal X-ray bursts, yet lower than those classified as intermediately long \citep[][]{chenevez2008,linares2012}. There are several possible explanations that may account for the unusual burst properties. 

First, for some LMXBs the first X-ray burst that occurred during a new outburst was found to be significantly longer than later bursts \citep[e.g., Aql X-1 and Cen X-4;][]{fushiki1992,kuulkers2008}. It is thought that at the start of the outburst the neutron star is relatively cold and therefore a thicker layer of fuel can build up before the ignition conditions are met. Since the duration of an X-ray burst depends on the cooling time of the ignition layer, a thicker layer would result in a longer (and more energetic) X-ray burst. By considering the expected burst recurrence time, we can assess whether this scenario might be applicable to J1850 and J1922.

With the burst energetics at hand, the ignition depth can be estimated as $y=E_b (1+z)/4\pi R^2 Q_{\mathrm{nuc,burst}}$, where $z$ is gravitational redshift, $R$ is the neutron star radius and $Q_{\mathrm{nuc,burst}}=1.6+4X~\mathrm{MeV~nucleon}^{-1}$, the nuclear energy release during an X-ray burst given an H-fraction $X$ at ignition \citep[e.g.,][]{galloway06}. For a neutron star with $M=1.4~\Msun$ and $R=10$~km (i.e., $z=0.31$), we estimate $y\simeq 10^{-8}~\mathrm{g~cm}^{-2}$ for the bursts of J1850 and J1922. 
The recurrence time that corresponds to a given ignition depth is $t_{\mathrm{rec}} \simeq y(1+z)/\dot{m}$, where $\dot{m}$ is the accretion rate onto the neutron star surface per unit area. For J1850 and J1922 we infer an average accretion rate during outburst of $\dot{m}=\dot{M}/4\pi R^2 \simeq300$ and $\simeq2300~\mathrm{g~cm}^{-2}~\mathrm{s}^{-1}$, respectively (assuming that the emission is isotropic). This would imply an expected recurrence time on the order of a few days to 2 weeks. 

The BAT transient monitor revealed activity from J1850 $\simeq5$ weeks before the BAT trigger of 2011 June 24. J1922 had been accreting for $\simeq3$~months prior to its BAT trigger on 2011 November 3, as evidenced by \swift\ and \maxi\ (Sections~\ref{subsec:new_intro} and~\ref{subsec:source_intro}). Comparing this to the expected burst recurrence time of $\lesssim2$ weeks suggests that the BAT triggers were likely not the first X-ray bursts that occurred during the outbursts of J1850 and J1922. The BAT coverage starts at 15~keV, which implies that it is not optimally sensitive to detect events as soft as X-ray bursts. These are therefore easily missed.

A second scenario that is worth considering is that some X-ray bursts display prolonged tails that last up to an hour \citep[e.g., EXO 0748--676 and the ``clocked burster'' GS 1826--24;][]{zand09}. These are explained as the cooling of layers below the ignition layer, that were heated by inward conduction of energy generated during the X-ray burst. These long cooling tails are characterized by fluxes and fluences that are two orders of magnitude lower than the prompt burst emission. In the case of J1850 and J1922, however, the fluence in the BAT peak and XRT tail are of similar magnitude. This indicates that the long duration of these X-ray bursts is likely caused by a different mechanism. 

Alternatively, the long burst duration may be explained in terms of a relatively large H-content in the ignition layer. The presence of H lengthens the duration of nuclear energy generation via the rp-process \citep[][]{schatz2001}. There are two additional effects that likely contribute in making the bursts observable for a longer time. At low accretion rates the temperature in the neutron star envelop is expected to be low compared to brighter bursters. This implies that a thicker fuel layer can accumulate before reaching the critical ignition conditions, resulting in a longer burst. Furthermore, J1850 and J1922 accreted at a relatively low level of $\simeq0.5\%-1\%$ of Eddington.
For such low persistent emission levels, the tails of the X-ray bursts are visible for a longer time (if observed with a sensitive instrument), which can add to a longer burst duration. The burst recurrence time at the accretion rates inferred for J1850 and J1922 is considerably lower than for higher rates. This can account for the fact that most observed X-ray bursts are shorter than those seen for these two sources. 

Intermediately long X-ray bursts are typically observed from sources accreting at $\simeq0.1\%-1\%$ of Eddington. These events are both longer ($\gtrsim10$~minutes) and more energetic ($\simeq10^{40-41}$~erg) than we have observed for J1850 and J1922 \citep[e.g.,][]{zand08,falanga09,linares09,degenaar2011_burst}. Some of these sources are strong candidate ultra-compact X-ray binaries. These contain an H-depleted companion star so that the neutron star accretes (nearly) pure He \citep[e.g.,][]{zand08}. In absence of H-burning, the temperature in the accreted envelop is relatively low, so that a thick layer of He can build up before it eventually ignites in a long and energetic X-ray burst. 
A similarly long X-ray burst has also been observed, however, from a source that accretes at $\simeq0.1$ of Eddington and shows a strong H-emission line in its optical spectrum. This testifies to the presence of an H-rich companion \citep[][]{degenaar2010_burst}. In this case it may be that unstable H-burning could not immediately trigger He ignition, allowing the development of a thick layer of fuel \citep[][]{cooper07,peng2007}. 

Summarizing, the burst duration and energetics of the X-ray bursts observed from J1850 and J1922 appear to fall in between that of normal and intermediately long X-ray bursts \citep[cf.][]{chenevez2008,linares2012}. Similar bursts have also been observed from the persistent neutron star LMXB 4U 0614+09 \citep[][]{kuulkers09,linares2012}. We considered several possible scenarios that may account for the relatively long burst duration, and find it most likely that it is due to the ignition of a relatively thick layer of (H-rich) fuel. Our findings suggests that there may not exist two distinct groups of X-rays bursts, but rather a continuous range of burst durations and energies. This supports the idea proposed by \citet{linares2012} that the apparent bimodal distribution is likely due to an observational bias toward detecting the longest and most energetic X-ray bursts from slowly accreting neutron stars.

\subsection{Quiescent Properties of J1850 and J1922}\label{subsec:q}
J1850 could not be detected in archival \xmm\ observations with an estimated upper limit of $L_q \lesssim(0.5-3.0)\times10^{32}~(D/3.7~\mathrm{kpc})^2~\lum$. J1922 is detected in quiescence with \swift/XRT and \suzaku\ at a 0.5--10 keV luminosity of $L_q \simeq (0.4-1.0) \times10^{32}~(D/4.8~\mathrm{kpc})^2~\lum$. These quiescent levels are common for neutron star LMXBs \citep[e.g.,][]{menou99,garcia01,jonker2004}.

It is thought that the accretion of matter onto the surface of a neutron star compresses the stellar crust and induces a chain of nuclear reactions that deposit heat \citep[][]{haensel1990a}. This heat spreads over the entire stellar body via thermal conduction and maintains the neutron star at a temperature that is set by the long-term averaged accretion rate of the binary \citep[][]{brown1998,colpi2001}. 

During quiescent episodes, the neutron star is  expected to thermally emit X-rays providing a candescent luminosity of $L_{\mathrm{q,bol}} = \langle \dot{M} \rangle Q_{\mathrm{nuc}} / m_u$, where $Q_{\mathrm{nuc}} \simeq 2$ MeV is nuclear energy deposited in the crust per accreted baryon (\citealt{haensel2008}; \citealt{gupta07}, but see \citealt{degenaar2011_terzan5_3}), $m_u=1.66\times10^{-24}$~g is the atomic mass unit (i.e., $Q_{\mathrm{nuc}} / m_u \simeq 9.6 \times 10^{17}~\mathrm{erg~g}^{-1}$), and $\langle \dot{M} \rangle$ is the long-term accretion rate of the binary averaged over $\simeq10^4$~yr (i.e., including both outburst and quiescent episodes). The latter can be estimated by multiplying the average accretion rate observed during outburst ($\dot{M}_{ob}$) with the duty cycle of the binary (i.e., the ratio of the outburst duration and recurrence time).

The outburst of J1850 observed with \swift\ in 2011 had a duration of $\simeq8$~weeks (0.17~yr) and an estimated bolometric accretion luminosity of $L_{\mathrm{bol}}\simeq7\times10^{35}~(D/3.7~\mathrm{kpc})^2~\lum$ (corresponding to $\dot{M}_{\mathrm{ob}}\simeq6\times10^{-11}~\mdot$). The outburst duration and recurrence time are not known for J1850, but we can make an order of magnitude estimate based on the constraints of the quiescent luminosity. 

Within the deep crustal heating model, a quiescent luminosity of $\lesssim3\times10^{32}~(D/3.7~\mathrm{kpc})^2~\lum$ would suggest a time-averaged mass-accretion rate of $\langle \dot{M} \rangle \lesssim 5 \times 10^{-12}~\mdot$. For $\dot{M}_{\mathrm{ob}}\simeq6\times10^{-11}~\mdot$, the corresponding duty cycle is $\lesssim10\%$. If J1850  typically exhibits outbursts with a duration of 8 weeks, the expected recurrence time would be $\gtrsim 2$~yr. We regard this as a lower limit, since the expected recurrence time increases for a longer outburst duration, or a quiescent thermal luminosity that is lower than the assumed upper limit of $3\times10^{32}~\lum$. This estimate is consistent with the fact that the \swift/BAT hard transient monitor did not detect any other outbursts from J1850 back to 2005 February \citep[][]{krimm2011}.

Since two outbursts with a relatively well-constrained duration have been observed for J1922, we can reverse the above reasoning and estimate the quiescent luminosity that is expected based on the observed duty cycle. The source was discovered when it exhibited an outburst of $\simeq20$~months in 2005--2006. Renewed activity was observed from the source 5 yr later in 2011--2012, when it accreted for $\simeq8-10$~months. For both outbursts, we estimate a similar bolometric accretion luminosity of $L_{\mathrm{bol}}\simeq(2-3)\times10^{36}~\lum$, which suggests a mean mass-accretion rate of $\dot{M}_{\mathrm{ob}}\simeq3\times10^{-10}~\mdot$ (Table~\ref{tab:spec}). If we assume a typical outburst duration of $\simeq1.3$~yr and a recurrence time of $\simeq5$~yr (i.e., a duty cycle of $\simeq$26\%), we can estimate a long-term averaged mass-accretion rate of $\langle \dot{M} \rangle \simeq 1\times10^{-10}~\mdot$ (equivalent to $\simeq6\times10^{15}~\mathrm{g~s}^{-1}$).

If the outburst behavior observed over the past decade is typical for the long-term accretion history of J1922, then the estimated $\langle \dot{M} \rangle$ should give rise to $L_{\mathrm{q,bol}} \simeq 6 \times10^{33}~(D/4.8~\mathrm{kpc})^2~\lum$. This is considerably higher than the observed $L_q \simeq 1 \times10^{32}~(D/4.8~\mathrm{kpc})^2~\lum$ (0.5--10 keV). Although the bolometric luminosity may be a factor of a few higher than that measured in the 0.5--10 keV band, the luminosity remains lower than expected based on the crustal heating model. Limited by a low number of counts, we cannot constrain the shape of the quiescent spectrum of J1922 with the current available data. It is possible that the quiescent emission contains a significant non-thermal component \citep[e.g.,][]{campana2005_amxps,wijnands05_amxps,heinke2009,cackett2010_cenx4, degenaar2012_1745,degenaar2012_amxp}, which would increase the discrepancy between the expected and observed quiescent thermal emission.

Provided that the heating models are correct, a plausible explanation for this apparent mismatch could be that the time-averaged mass-accretion rate is lower than estimated (e.g., because the recent outburst behavior is not representative for the long-term accretion history). Alternatively, the neutron star could be cooling faster than assumed in the standard paradigm \citep[][]{page2004}. A comparison with theoretical cooling models of \citet{yakovlev2004} would suggest enhanced cooling due to the presence of kaons in the neutron star core. 

Four neutron star LMXBs have been closely monitored after the cessation of their very long ($\gtrsim1$~yr) accretion outbursts, which has revealed that the neutron star temperature was gradually decreasing over the course of several years \citep[][]{wijnands2001,wijnands2003,cackett2008,cackett2010,degenaar09_exo1,degenaar2010_exo2,fridriksson2010,fridriksson2011,diaztrigo2011}. Recently, similar behavior has been observed for a transient neutron star LMXB in Terzan 5 that exhibited a much shorter accretion outburst of $\simeq10$~weeks \citep[][]{degenaar2011_terzan5_2,degenaar2011_terzan5_3}. These observations can be explained as cooling of the neutron star crust, which became considerably heated during the accretion outburst and needs time to cool down in quiescence. Monitoring and modeling this crustal cooling provides the unique opportunity to gain insight into the properties of the neutron star crust and core \citep[][]{rutledge2002,wijnands04_quasip,shternin07,brown08,degenaar2011_terzan5_3,page2012}. 

With its relatively low quiescent luminosity, long outburst duration and low extinction, J1922 would be a promising target to search for crustal cooling now that its recent outburst has ceased.

\acknowledgments
N.D. is supported by NASA through Hubble Postdoctoral Fellowship grant number HST-HF-51287.01-A from the Space Telescope Science Institute, which is operated by the Association of Universities for Research in Astronomy, Incorporated, under NASA contract NAS5-26555. R.W. is supported by a European Research Council (ERC) starting grant, and M.L. by a Rubicon fellowship from the Netherlands Organization for Scientific Research (NWO). N.D. thanks Min-Su Shin for useful discussion about flaring M-stars. This work made use of data supplied by the UK Swift Science Data Centre at the University of Leicester and the public data archives of \swift, \xmm\ and \suzaku. 

{\it Facilities:} \facility{\swift\ (BAT/XRT/UVOT)}, \facility{{\it XMM} (EPIC)}, \facility{\suzaku\ (XIS)}

\appendix

\setcounter{figure}{0}
\renewcommand{\thefigure}{A.\arabic{figure}}

\setcounter{table}{0}
\renewcommand{\thetable}{A.\arabic{table}}

\section{Serendipitous faint X-ray sources}\label{appendix}
The XRT images of the field around J1922 reveal two additional faint X-ray sources (Figure~\ref{fig:ds9_1922}). Neither of these objects are cataloged in the SIMBAD data base. We assigned them \swift\ names and determined their positions and basic properties (Table~\ref{tab:others}). We used the tools \textsc{xrtcentroid} and \textsc{uvotcentroid} to determine their positions. 

Source \#1 is located in the wing of the point spread function of J1922, and is most clearly visible when the transient is in quiescence (see Figure~\ref{fig:ds9_1922}). To avoid contamination, we extracted information for source \#1 using only observations 35471004--7 (when J1922 was not active; Table~\ref{tab:obs}). The object is detected at an average XRT count rate of $\simeq 1\times10^{-3}~\cnts$ (0.5--10 keV). There are not enough counts collected to extract an X-ray spectrum. DSS and UVOT images reveal a possible counterpart at the XRT position (Table~\ref{tab:others}). This object is detected in some of the UVOT images (at $3.3-12.2\sigma$ significance) with $v\simeq19.4-20.3$~mag. We derive a position of R.A. = 19:22:39.61, decl. = -17:18:08.3 (J2000), with a 90\% confidence uncertainty of $0.9''$. For other observations we obtain upper limits of $v>18.8-20.2$~mag. The intensity is close to the detection limit of the UVOT observations.

\begin{figure*}
 \begin{center}
\includegraphics[width=5.8cm]{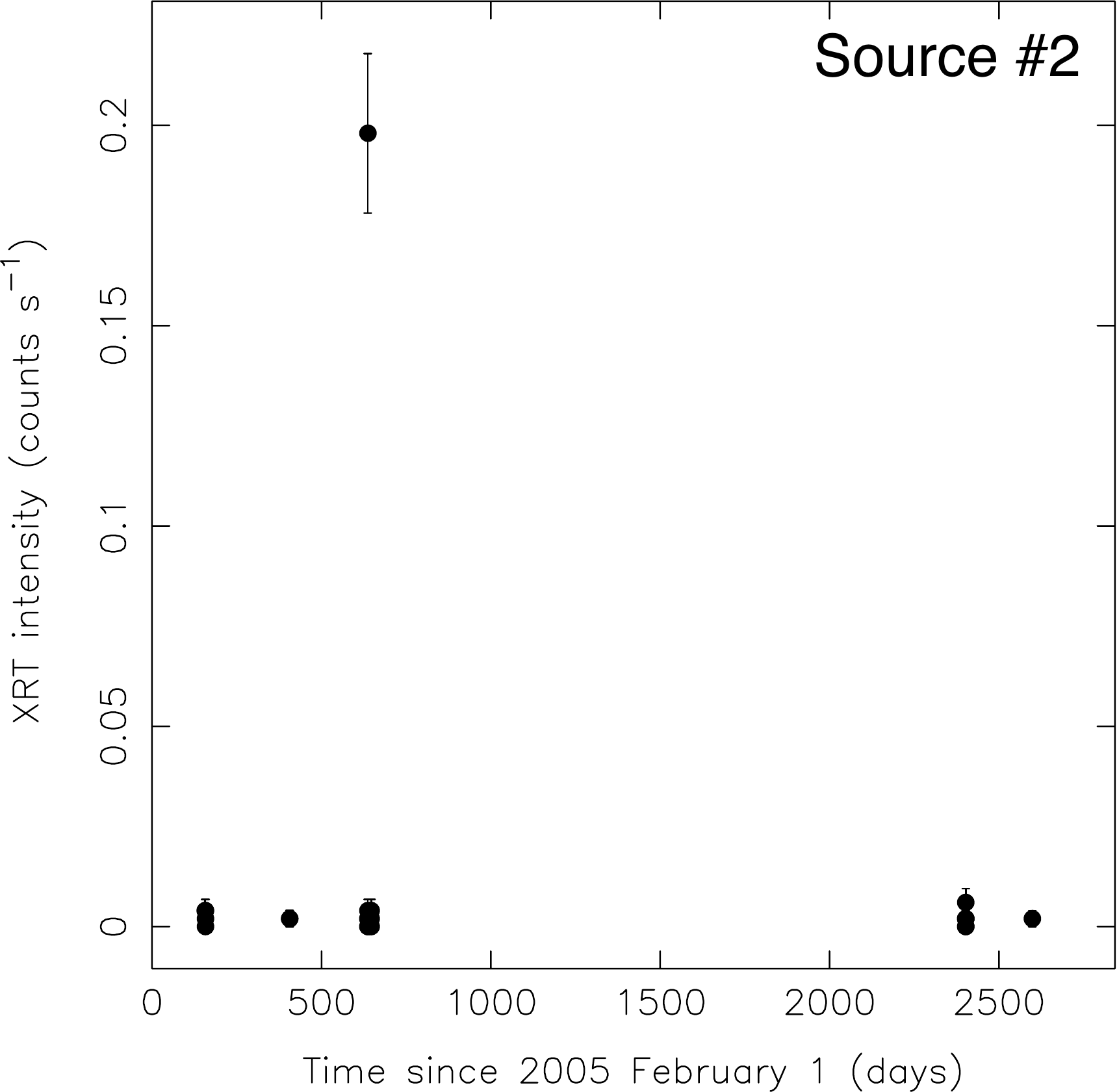}\hspace{0.2cm}
\includegraphics[width=5.8cm]{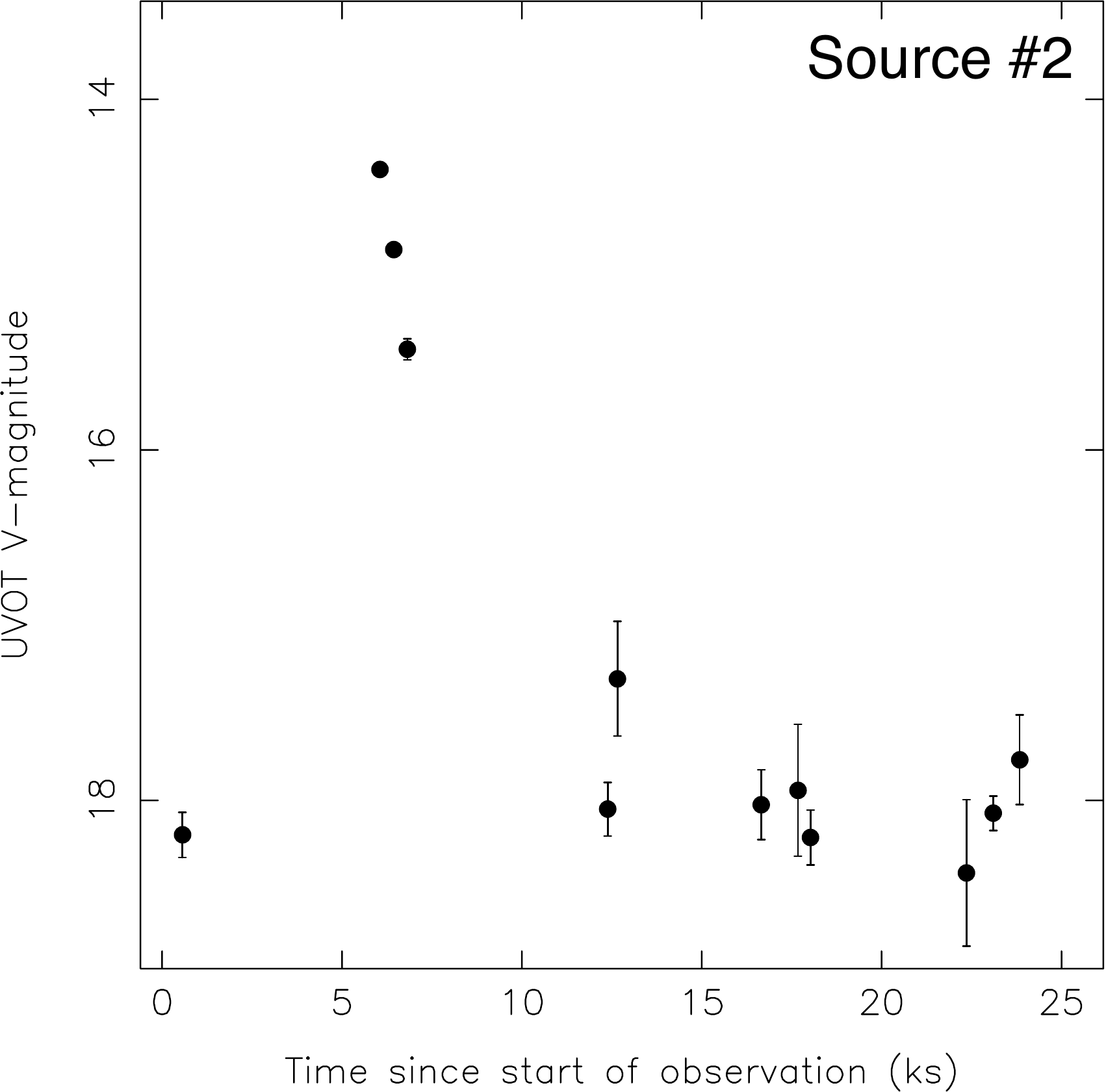}\hspace{0.2cm}
\includegraphics[width=5.8cm]{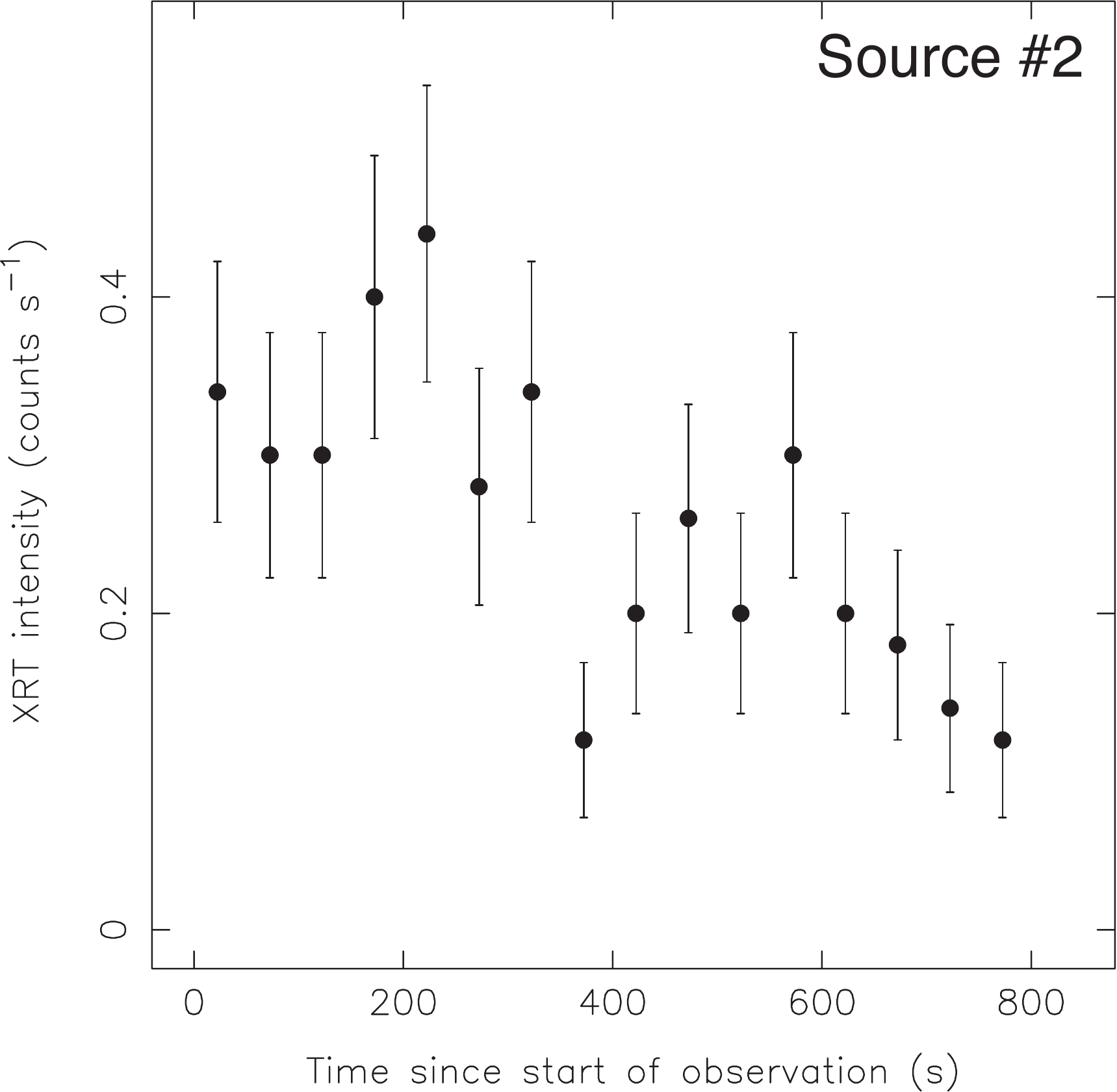}
    \end{center}
\caption[]{{Results for source \#2 (Swift J192241.1--272040), that was serendipitously detected within FOV of J1922. Left: XRT light curve obtained between 2005 and 2012 (binned per orbit). Middle: evolution of the UVOT $v$-band magnitude during the flare observation of 2006 October 30. Right: XRT light curve of the orbit containing the flare (50 s resolution).}}
 \label{fig:other}
\end{figure*}

Source \#2 is detected at an average XRT count rate of $\simeq2\times10^{-3}~\cnts$, but experienced an X-ray flare that was a factor $\simeq100$ brighter on 2006 October 30 (Obs ID 35471004; Figure~\ref{fig:other} left). This observation had a total exposure time of $\simeq5.2$~ks and consisted of five separate orbits. The enhanced activity is detected in the second orbit, which had an exposure time of $\simeq850$~s. During this interval, the source intensity decreased from $\simeq0.4$ to $\simeq0.1~\cnts$, but remained well above $\simeq10^{-3}~\cnts$ (Figure~\ref{fig:other} right). In both the preceding and subsequent orbits, the source is detected at its persistent count rate, which implies that the X-ray flare had a total duration between 0.25 and 3 hr.

\begin{figure}
 \begin{center}
\includegraphics[width=8.0cm]{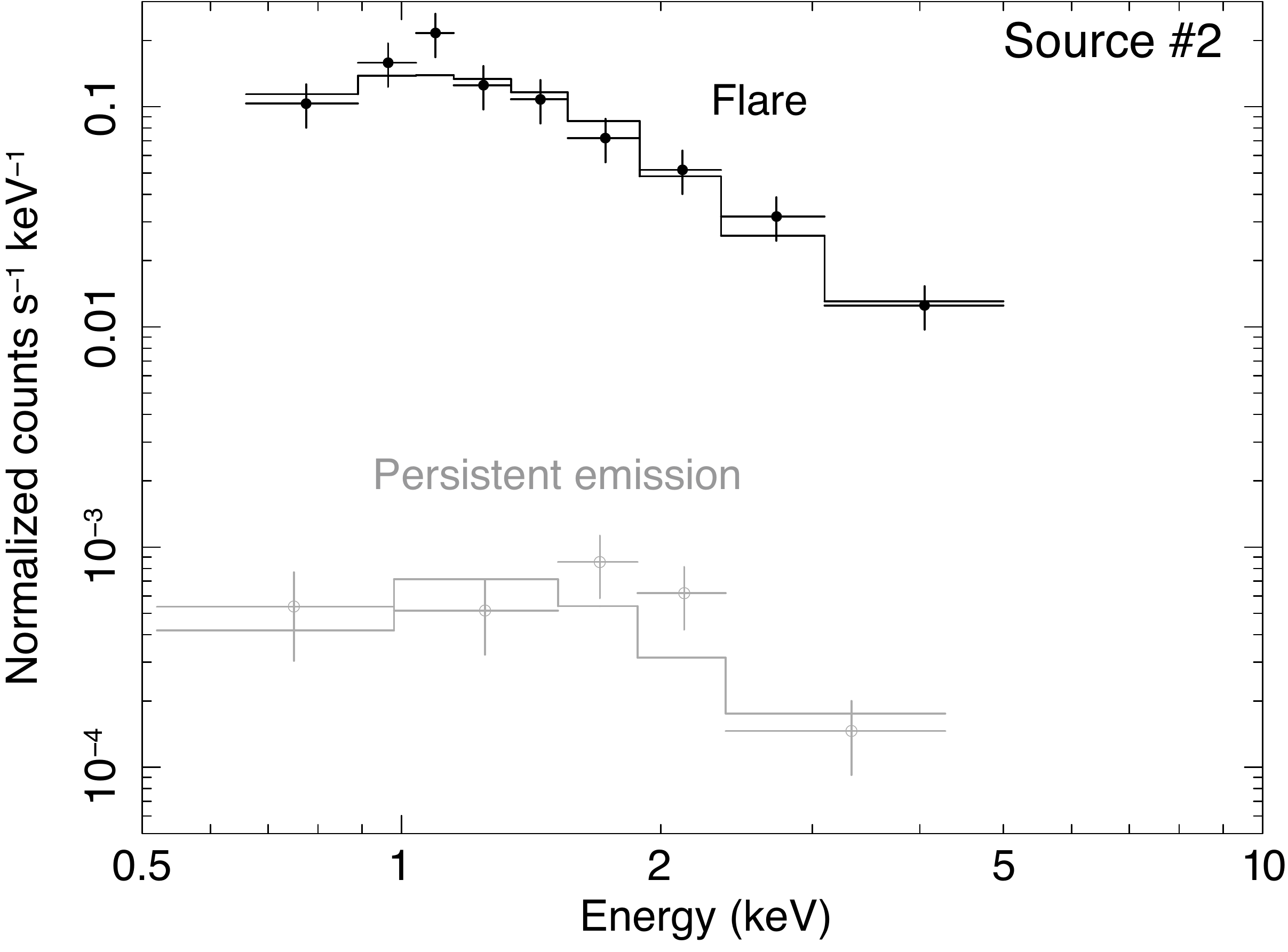}
    \end{center}
\caption[]{{\swift/XRT spectra of source \#2 for the flare (black) and persistent (gray) emission, together with the best-fit absorbed power-law model. 
}}
 \label{fig:other2}
\end{figure}

We extracted X-ray spectra of both the flare and the persistent emission. We fitted these simultaneously to an absorbed power-law model with the hydrogen column density tied between the two data sets. This resulted in $N_{\mathrm{H}}=(1.8\pm0.2)\times10^{21}~\nh$ and $\chi^2_{\nu}=1.1$ for 9 dof (Figure~\ref{fig:other2}). The persistent X-ray spectrum can be described with $\Gamma = 1.5\pm0.7$, while the flare appears to be softer with $\Gamma=2.1\pm0.5$. The unabsorbed 0.5--10 keV fluxes are $(1.2\pm 0.3)\times10^{-13}$ and $(1.7\pm 0.5)\times10^{-11}~\flux$ for the persistent and flare emission, respectively. 

DSS and UVOT images reveal a possible counterpart centered at the XRT position of source \#2 (Figure~\ref{fig:ds9_1922}). We find magnitudes of $v\simeq 17.6-18.2$~mag for the persistent emission of this object, while it brightened up to $v\simeq 14.4$~mag during the second orbit of the flare observation. Figure~\ref{fig:other} (middle) displays the evolution of the UVOT $v$-magnitude during that observation. The simultaneous brightening in the optical and X-ray bands firmly establishes that the object seen in the UVOT images is indeed the counterpart to source \#2. We obtained a sub-arcsecond localization from the UVOT data (Table~\ref{tab:others}).

There are two observations in which source \#2 is detected in both the $b$ and the $v$ bands, which allows for a color determination. This gives $(b-v)=1.36-1.46$~mag. The source is not detected with the $u$-filter on these occasions, which implies $(u-b)>0.9$. Comparing these colors and the observed $v$-magnitude to stars of different spectral classes \citep[][]{drilling2000}, we find that the UVOT object could be an M-dwarf located at a distance of $D\simeq0.4$~kpc \citep[where we have taken into account the visual extinction in the direction of the source of $A_V\simeq1$~mag, as determined using the relation of][]{guver2009}. 

M-dwarfs are the primary stellar component in the Galaxy by number \citep[][]{bochanski2010} and are known to exhibit flares that are visible across the electromagnetic spectrum \citep[e.g.,][]{hawley2003}. At a distance of 0.4~kpc, the observed persistent emission and that of the flare translate into luminosities of $\simeq2\times10^{30}$ and $~\simeq2\times10^{32}~\lum$, respectively. These intensities, as well as the observed duration and amplitude of the flare, are consistent with the properties of flaring M-dwarfs \citep[e.g.,][]{hawley2003}. We therefore tentatively classify source \#2, Swift J192241.1--272040, as a nearby Galactic M-dwarf.\\

\begin{table*}[h!]
\begin{center}
\caption{Source Properties.}
\begin{threeparttable}
\begin{tabular}{c c c c c c c}
\toprule
Source \# & Name & R.A. & Dec. & Error & XRT Count Rate & UVOT $v$-magnitude \\
 & & (hh mm ss.ss) & ($^{\circ}~'~''$) & ($''$) & ($10^{-3}~\cnts$)  & (mag) \\
\midrule
1 & Swift J192239.5--171808 & 19 22 39.54 & --17 18 08.0 & 3.8 & $\simeq1$ & 19.4--20.3 \\
2 & Swift J192241.1--272040 & 19 22 41.15 & --17 20 39.4 & 0.7 & $\simeq2-250$ & 14.4--18.2  \\
\bottomrule
\end{tabular}
\label{tab:others}
\begin{tablenotes}
\item[]Note. -- The listed coordinates refer to the J2000 epoch. Errors represent a 90\% confidence level. The XRT count rates are for the 0.5--10 keV energy band.
\end{tablenotes}
\end{threeparttable}
\end{center}
\end{table*}

\end{document}